\documentclass[ reprint,
showpacs,  
aps,
prl,
floatfix,superscriptaddress,
]{revtex4-1}
\usepackage{amsmath,amsfonts,amsthm}
\usepackage{dsfont}
\usepackage{braket}
\usepackage{graphicx}
\usepackage{epstopdf}
\usepackage{dcolumn}
\usepackage{bm}
\usepackage{float}
\usepackage[caption=false]{subfig}
\usepackage{color}
\usepackage{float}
\usepackage{hyperref}

\begin{document}

\preprint{APS/123-QED}

\title{Detection of quantum phases via out-of-time-order correlators}
\author{Ceren B. Da\u{g}}
\email{cbdag@umich.edu}
\author{Kai Sun}
\affiliation{Department of Physics, University of Michigan, Ann Arbor, Michigan 48109, USA}
\author{L.-M. Duan}
\affiliation{Center for Quantum Information, IIIS, Tsinghua University, Beijing 100084, PR China.}
\date{\today}

\begin{abstract}
We elucidate the relation between out-of-time-order correlators (OTOCs) and quantum phase transitions via analytically studying the OTOC dynamics in a degenerate spectrum. Our method points to key ingredients to dynamically detect quantum phases via out-of-time-order correlators for a wide range of quantum phase transitions and explains the existing numerical results in the literature. We apply our method to a critical model, XXZ model that numerically confirms our predictions.
\end{abstract}

\maketitle

%\tableofcontents
%%%%%%%%%%%%%%%%%%%%%%%%%%%%%%%%%%%%%%%%%%%%%%%%%%%%%%%%%%%%%%%%%%%%%%%%%%%

Out-of-time-order correlators (OTOCs) \cite{1969JETP...28.1200L} probe information scrambling in quantum systems of different nature \cite{2008JHEP...10..065S,2013JHEP...04..022L,2017PhRvB..95f0201S,2017JHEP...10..138H,articleRey,doi:10.1002/andp.201600332,2017NJPh...19f3001B,2018arXiv180711085D} and reflect the symmetries \cite{doi:10.1002/andp.201600332,2017PhRvB..95f0201S,2018arXiv180711085D} or lack thereof \cite{2008JHEP...10..065S,Maldacena2016,doi:10.1002/andp.201600332} of the underlying Hamiltonian. An OTOC, unlike a time-ordered four-point (or two-point) correlator \cite{doi:10.1002/andp.201600318}, can determine the spatial and temporal correlations throughout the system, thus giving rise to a bound on information spread \cite{PhysRevB.96.020406,2018arXiv180200801X,2018arXiv180711085D}. Through such bounds and the decay rate of an OTOC, one can dynamically detect thermal \cite{doi:10.1002/andp.201600332,2017NJPh...19f3001B,2018arXiv180711085D} and localized phases \cite{doi:10.1002/andp.201600332,2017PhRvB..95f0201S,doi:10.1002/andp.201600318,2016arXiv160801914F,2017PhRvB..95e4201H,2018arXiv180711085D}. Recently OTOC has been numerically observed to be susceptible to phase transitions either signaling criticality in a diverging Lyapunov exponent \cite{2017PhRvB..96e4503S} or showing signatures of symmetry-broken phases in its saturation value \cite{PhysRevLett.121.016801}. The latter led to more research that shows the relation emerging in other forms, e.g. in excited states \cite{2018arXiv181201920W}, or with more experimentally-relevant platforms and system parameters \cite{2018arXiv181111191S}. The interest in providing more verification for such an emergent relation is understandable, not only because the relation points to a practical potential for OTOC in dynamically probing quantum criticality, but also the relation is received as unexpected \cite{PhysRevLett.121.016801}. It is indeed an intriguing question how a chaos-detecting and out-of-time ordered correlator that is contributed by presumably all spectrum could also probe ground state physics. The reasons of this relation remain unknown as well as an answer to whether the relation is universal. Motivated by these questions, here, we develop a method on OTOC dynamics to obtain intuition for the emerging relation between quantum phase transitions and out-of-time-order correlators. Remarkably it is possible to dynamically decompose OTOC and show that the ground state physics is the leading order contribution to it under the criteria that our method provides. This is the origin why OTOC saturation value could detect the ground state degeneracy. Therefore, we reach to the conclusion that the OTOC is susceptible to long-range order, while the quasi-long range order is not visible to it. Our method provides additional insights regarding the relation, e.g. (i) the relation is not restricted to already-studied models and 1D \cite{PhysRevLett.121.016801,2018arXiv181111191S}; (ii) the relation can be extended to include the phase transitions in other eigenstates \cite{2018arXiv181201920W}. Hence, our theory elucidates the reasons of this \emph{unexpected} connection, renders it intuitive and universal with further insights. To verify our method, we study the dynamics of 1D critical XXZ chain, where there are Ising and critical XY phases.

\emph{Method.} Our aim is to be able to come up with an expression that predicts the saturation value of OTOC for long times in the spirit of Eigenstate Thermalization Hypothesis (ETH) \cite{1999JPhA...32.1163S, 1995chao.dyn.11001S}. The out-of-time-order correlation function can be defined as
\begin{eqnarray}
F(t) &=& \left\langle W^{\dag}(t) V^{\dag} W(t) V \right\rangle,
\label{OTOCEq}
\end{eqnarray}
where $V$ and $W$ are local operators and the expectation value is over an initial state $\Ket{\psi(0)}$. This initial state could be chosen as the ground state \cite{PhysRevLett.121.016801,articleRey}, or a random Haar-distributed state \cite{PhysRevB.96.020406,2018arXiv180711085D} to approximate an equiprobable state $\mathcal{I}$ in Eq.~\eqref{OTOCEq} \cite{PhysRevLett.108.240401,2017AnP...52900350L,supp}. Eventually, the original definition that is the commutator growth $- \text{Tr}\left(\frac{\text{exp}\left[-\beta H \right]}{Z}\left[W(t),V \right]^2\right)$ \cite{Maldacena2016} could be reexpressed in terms of the OTOC of operators $W$ and $V$ with an initial state at the inverse temperature $\beta$. Therefore we can probe the information scrambling through OTOCs \cite{2017PhRvX...7c1011L,articleRey,PhysRevA.94.040302,2017NJPh...19f3001B,2018arXiv180711085D}. 

Given $\Ket{\psi(t)} = \sum_{\alpha} c_{\alpha} e^{-i E_{\alpha}t} \Ket{\psi_{\alpha}}$, where $\Ket{\psi_{\alpha}}$ are eigenstates of the Hamiltonian with the associated eigenvalues $E_{\alpha}$, we define a modified initial state $\Ket{\psi'(0)} = V \Ket{\psi(0)}$ and have $\Ket{\psi'(t)} = \sum_{\beta} b_{\beta} e^{-i E_{\beta}t} \Ket{\psi_{\beta}}$. Then the OTOC, Eq.~\eqref{OTOCEq}, can be recast to a fidelity measure of 3-point function, and with the help of completeness relation $\sum_{\gamma} \Ket{\psi_{\gamma}}\Bra{\psi_{\gamma}} = \mathbb{I}$ becomes
\begin{eqnarray}
F(t) = \sum_{\alpha,\beta,\gamma,\gamma'} c_{\alpha}^* b_{\beta} e^{-i (E_{\beta}-E_{\alpha} + E_{\gamma}-E_{\gamma'})t} W^{\dagger}_{\alpha \gamma} V^{\dagger}_{\gamma \gamma'} W_{\gamma' \beta},\notag
%\label{OTOCdynamics}
\end{eqnarray}
where $\Bra{\psi_{\alpha}} W \Ket{\psi_{\gamma}} = W_{\alpha \gamma}$ are eigenstate expectation values \cite{2008Natur.452..854R}. Now one can derive the saturation value for long times as well as dynamical features, such as revival timescales in integrable Hamiltonians \cite{PhysRevA.97.023603}. 

We study the saturation value in long times, since this value is expected to contain the signature of quantum phases. 
%Note that a similar formalism is known to give the dynamical features of expectation values evolved in time \cite{1995chao.dyn.11001S,PhysRevA.97.023603}, though, less involved than the expressions for OTOCs. 
For long enough times, equilibration in OTOC dynamics can be obtained only when the phase decoheres. Then the equilibration value can be obtained by requesting $E_{\beta}-E_{\alpha} + E_{\gamma}-E_{\gamma'}=0$. This condition can be satisfied with four different scenarios: (i) $E_{\alpha}=E_{\beta}$ and $E_{\gamma}=E_{\gamma'}$; (ii) $E_{\alpha}=E_{\gamma}$ and $E_{\beta}=E_{\gamma'}$; (iii) $E_\alpha=E_\beta=E_\gamma=E_{\gamma'}$, which is contained both in (i) and (ii); (iv) $E_{\beta}-E_{\alpha} + E_{\gamma}-E_{\gamma'}=0$ with $E_{\beta}\neq E_{\alpha} \neq E_{\gamma}\neq E_{\gamma'}$. If a nondegenerate spectrum is assumed, i.e. $E_{\alpha}=E_{\beta}$ implies $\alpha=\beta$, the OTOC reduces to,
\begin{eqnarray}
&F_{t\rightarrow \infty}& = \sum_{\alpha,\gamma} c_{\alpha}^* b_{\alpha} |W_{\alpha \gamma}|^2 V^{\dagger}_{\gamma \gamma} + \sum_{\alpha,\beta} c_{\alpha}^* b_{\beta} W^{\dagger}_{\alpha \alpha} V^{\dagger}_{\alpha \beta} W_{\beta \beta} \label{saturationEq}\\
&-& \sum_{\alpha} c_{\alpha}^* b_{\alpha} |W_{\alpha \alpha}|^2 V^{\dagger}_{\alpha \alpha} + \sum_{\alpha \neq \beta \neq \gamma \neq \gamma'} c_{\alpha}^* b_{\beta} W^{\dagger}_{\alpha \gamma} V^{\dagger}_{\gamma \gamma'} W_{ \gamma' \beta}\notag,
\end{eqnarray}
with four terms corresponding to four conditions (i)-(iv), respectively. We note that writing OTOC as in Eq.~\eqref{saturationEq} proved to be useful previously to understand the quantum chaotic systems better, e.g. in chaotic spin chains with conserved quantities that also obey ETH, decay to zero is not supposed to be exponential, but inverse polynomial in system size \cite{PhysRevLett.123.010601} and OTOCs capture eigenstate correlations that ETH cannot \cite{PhysRevLett.122.220601}. These correlations can readily be seen in the first, second and the fourth terms of Eq.~\eqref{saturationEq}. See Supplement S5 \cite{supp} for some remarks that immediately follow from Eq.~\eqref{saturationEq} about systems with nondegenerate chaotic spectra. Now we are going to generalize Eq.~\eqref{saturationEq} to a more generic form, which allows degeneracy in the energy spectra, because a quantum phase transition usually involves energy degeneracy, e.g. degeneracy from spontaneous symmetry breaking or other sources \cite{2019arXiv190605241D}. We group all eigenstates of the Hamiltonian into degenerate sets labeled by $\theta$, and each state in its corresponding set is denoted by $\alpha$ for an eigenstate $\psi_{[\theta,\alpha]}$. The OTOC can be reorganized with the new notation, which is one main result of this Letter,
\begin{widetext}
\begin{eqnarray}
&F(t\rightarrow \infty)& = \sum_{\theta\theta'} \sum_{\alpha \beta \gamma \gamma'}c^*_{[\theta,\alpha]} \left(  b_{[\theta,\beta]} W^{\dagger}_{[\theta,\alpha][\theta',\gamma]}V^{\dagger}_{[\theta',\gamma][\theta',\gamma']}W_{[\theta',\gamma'][\theta,\beta]} + b_{[\theta',\beta]} W^{\dagger}_{[\theta,\alpha][\theta,\gamma]}V^{\dagger}_{[\theta,\gamma][\theta',\gamma']}W_{[\theta',\gamma'][\theta',\beta]} \right) \label{DegsaturationEq} \\ 
&+ \sum_{\alpha \beta \gamma \gamma'}& \left( -\sum_{\theta} c^*_{[\theta,\alpha]} b_{[\theta,\beta]} W^{\dagger}_{[\theta,\alpha][\theta,\gamma]}V^{\dagger}_{[\theta,\gamma][\theta,\gamma']}W_{[\theta,\gamma'][\theta,\beta]} + \sum_{\theta\neq\theta' \neq \phi \neq \phi'} c^*_{[\theta,\alpha]} b_{[\theta',\beta]} W^{\dagger}_{[\theta,\alpha][\phi,\gamma]}V^{\dagger}_{[\phi,\gamma][\phi',\gamma']}W_{[\phi',\gamma'][\theta',\beta]} \right). \notag
\end{eqnarray}
\end{widetext}
Here, $\theta, \theta', \phi, \phi'$ denote degenerate sets while $\alpha,\beta,\gamma,\gamma'$ denote quantum states in their corresponding sets. Eq.~\eqref{DegsaturationEq} can predict the saturation value of OTOC accurately if the OTOC saturates at a finite time. If the OTOC does not saturate or shows transient effects, Eq.~\eqref{DegsaturationEq} still predicts the time-average of OTOC signal $\bar{F}=1/\mathcal{T} \int dt F(t)$ over a time interval $\mathcal{T}$ with sufficient accuracy. In this sense, Eq.~\eqref{DegsaturationEq} is not limited to long-time dynamics $t\rightarrow \infty$ \cite{supp}.

We look for the criteria that the ground state subspace contribution is leading order in the OTOC saturation value Eq.~\eqref{DegsaturationEq}. For this, we first set $W=V$ as the order parameter operator in Eq.~\eqref{DegsaturationEq} for convenience. Then we expand the coefficients $b_{[\theta,\beta]}=\sum_{\kappa,\tau}W_{[\theta,\beta][\kappa,\tau]}c_{[\kappa,\tau]}$ in Eq.~\eqref{DegsaturationEq} by using the initial state. If (i) the initial state is set to the state where the phase transition is expected to happen, e.g. the ground state(s) $c_{[1,1]}=1$; and (ii) we apply an ansatz on the matrix elements of the operator projected on this state, e.g. $|W_{[1,\alpha][\theta,\beta]}|^2 \ll 1$, where $\theta\neq 1$ is a different energy subspace than the subspace of the ground state(s), we observe the following dynamical decomposition:
\begin{eqnarray}
F(t\rightarrow \infty) &=& F_{\text{gs}}(t\rightarrow \infty) + F_{\text{ex}}(t\rightarrow \infty).
\end{eqnarray}
Here $F_{\text{gs}}(t\rightarrow \infty)$ is the ground subspace contribution, whereas the $F_{\text{ex}}(t\rightarrow \infty)$ is the contribution of higher energy excitations. The latter is a correction to the ground-state physics in the OTOC, when the criteria are satisfied. The assumption on the initial state sets the scrambling discussed in the rest of the paper to effectively zero temperature. Whereas the operator ansatz becomes even more specific for the phase of interest. If there is a symmetry-broken long-range order to capture, the fluctuations between the matrix elements of the operator are suppressed in the ground state subspace, meaning there is at least a pair of matrix elements accumulating the order $\rightarrow |W_{[1,\alpha][1,\beta]}|^2 \sim \mathcal{O}\left(1\right)$. This modifies the operator ansatz as $|W_{[1,\alpha][1,\beta]}|^2 \gg |W_{[1,\gamma][\theta,\gamma']}|^2$ for the ordered phase. Thus, we derive the expression for $F_{\text{gs}}(t\rightarrow \infty)$ in the ordered phase as,
\begin{eqnarray}
&F_{\text{gs}}(t\rightarrow \infty)& \sim \label{degSubs}\\
&\sum_{\beta,\gamma,\gamma'}& W_{[1,1][1,\gamma]}W_{[1,\gamma][1,\gamma']}W_{[1,\gamma'][1,\beta]}W_{[1,\beta][1,1]},\notag
\end{eqnarray}
while the operator ansatz simultaneously implies that the OTOC is dominated by the ground state physics, $F_{\text{gs}} \gg F_{\text{ex}}$ in the ordered phase. On the other hand, the fluctuations between the matrix elements of the operator are maximal in a disordered phase, implying $W_{[1,\alpha][1,\beta]} \sim 0$ for all in the ground state subspace which results in $F_{\text{gs}}(t\rightarrow \infty)\sim 0$. Therefore, the OTOC is dominated by the correction terms that are contributed by the excitations in the spectrum $F_{\text{ex}}(t\rightarrow \infty)$. This result is an important insight that originates from the dynamical decomposition method and cannot be observed only via real-time dynamics simulations, e.g. in Ref.~\cite{PhysRevLett.121.016801}. In addition, the operator ansatz $|W_{[1,\alpha][\theta,\beta]}| \ll 1$ guarantees a bounded correction term $F_{\text{ex}}(t\rightarrow \infty) \ll 1$. As a result, (i) the OTOC is able to capture the degeneracy in the ground state (Eq.~\eqref{degSubs}) and, (ii) the correction of the excited states always remains bounded; all of which explains why the OTOC differentiates an ordered phase from a disordered one, e.g. in ground state \cite{PhysRevLett.121.016801} or excited-state \cite{2018arXiv181201920W} phase transitions. A mixed initial state (e.g. finite or infinite temperature) violates the initial state assumption, hence suggesting a smoothed phase boundary by washing away the sharp signature at the transition point \cite{2018arXiv181111191S}. Hence the dynamical decomposition method reveals the key ingredients of the emergent relation between information scrambling and symmetry-breaking phase transitions, rendering this unexpected numerical observation \cite{PhysRevLett.121.016801} a fundamental connection.

Advanced numerical methods (Lanczos, tensor networks) can be employed to determine only the lowest-lying states to give the leading order term in OTOC, Eq.~\eqref{degSubs}. In this sense, Eq.~\eqref{degSubs} provides us a low-cost alternative to simulating the real-time OTOC dynamics in the computation of the OTOC saturation value when we use the OTOC to probe criticality. Finally, we predict that the ground state contribution to the OTOC saturation cannot efficiently distinguish quasi-long range order from a disordered phase. Because, the quasi-long range order produces zero expectation value for the order parameter (per site): $W_{[1,\alpha][1,\beta]} \sim 0$, similar to a disordered phase, and hence $F_{\text{gs}}(t\rightarrow \infty) \sim 0$ follows with correction term dominating the OTOC saturation $F(t\rightarrow \infty)$. In the following we will provide verification for our method and theory on the 1D XXZ model.

\begin{figure}
\centerline{\includegraphics[width=0.45\textwidth]{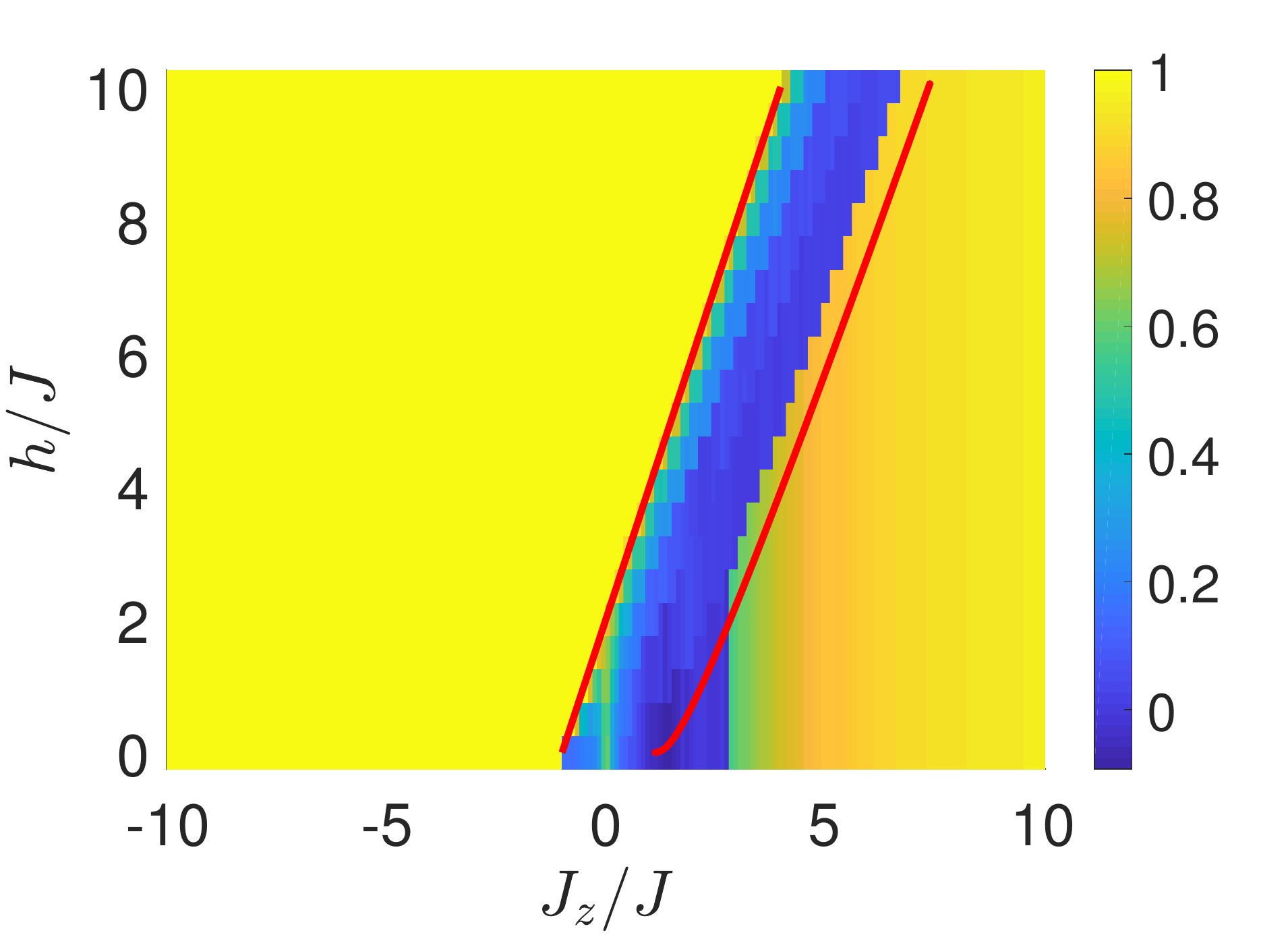}}
\caption{Phase diagram based on the OTOC saturation values via Eq.~\eqref{DegsaturationEq}, x-axis is the spin interaction strength in the z-direction $J_z$ and y-axis is the magnetic field $h$, for $N=14$ system size and $\sigma^z_i$ bulk spin operator, when periodic boundary conditions are set and the initial state is a ground state. The red lines are the phase boundaries based on Bethe ansatz technique for infinite-size system \cite{2017LNP...940.....F}.}
\label{Fig1}
\end{figure}

\emph{Model and results.} The Hamiltonian of the XXZ model reads,
\begin{eqnarray}
H&=&J \sum_i \left( \sigma_i^x \sigma_{i+1}^x + \sigma_i^y \sigma_{i+1}^y + \frac{J_z}{J} \sigma_i^z \sigma_{i+1}^z\right)  + h \sum_i \sigma^z_i,\notag
\end{eqnarray}
where $\sigma_i^n$ are spin-1/2 Pauli matrices with energy scale set to $J$ and hence time scale set to $1/J$; $J_z/J$ and $h$ are the z-axis spin coupling strength and the magnetic field strength, respectively. The red lines in Fig.~\ref{Fig1} show the phase boundaries produced by an exact method (Bethe Ansatz) for an infinite-size system. Therefore, this model has three phases: two gapped Ising phases (ferromagnetic and antiferromagnetic) at large $|J_z/J|$ and a gapless XY phase with quasi-long range order for small $|J_z/J|$, i.e. the Berezinskii-Kosterlitz-Thouless transition \cite{1971JETP...32..493B,1972JPhC....5L.124K}. We choose the OTOC operators as $\sigma^z_i$ or $\sigma^x_i$ for the spin at the $i$th site, based on the order parameters of the ferromagnetic Ising phase ($\sum_i \sigma^z_i$), antiferromagnetic Ising phase ($\sum_i (-1)^i \sigma^z_i$), and the XY-phase ($\sum_i \sigma^x_i$). Fig.~\ref{Fig1} shows the phase diagram based on the saturation values of OTOCs with $\sigma^z_i$ [computed using Eq.~\eqref{DegsaturationEq} for a system of $N=14$ spins]. We numerically confirm our theory with OTOC saturation values that are either nonzero or nearly zero in the Ising and XY phases, respectively. In fact, the OTOC recovers the phase boundaries of the Bethe ansatz solution: the agreement is perfect at the ferromagnetic-XY phase boundary and approximate at the antiferromagnetic-XY boundary due to significant finite-size effects \cite{supp}. 

\begin{figure}
\centering
\subfloat[]{\label{fig2a}\includegraphics[width=0.24\textwidth]{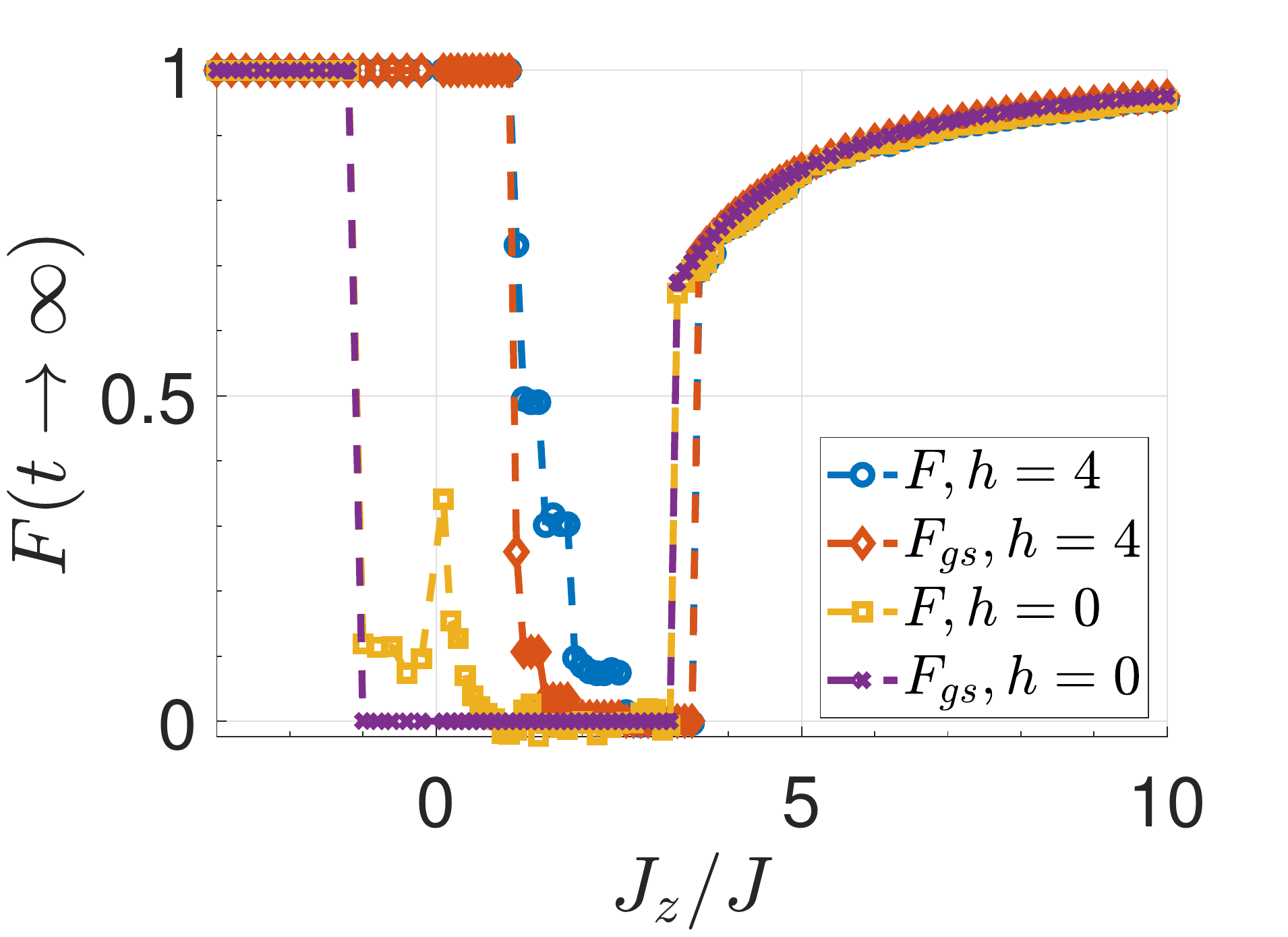}}\hfill 
\subfloat[]{\label{fig2b}\includegraphics[width=0.24\textwidth]{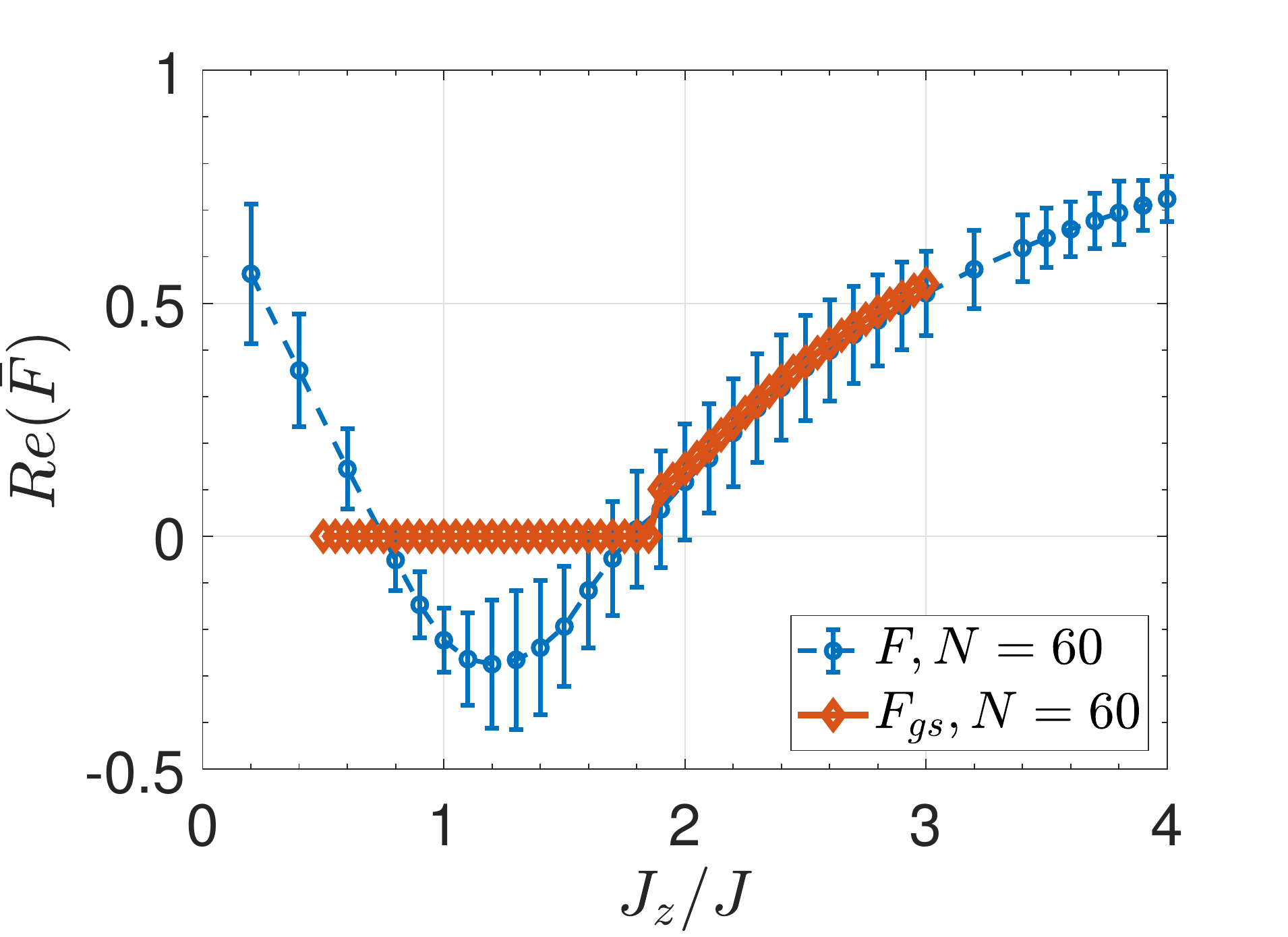}}\hfill 
\subfloat[]{\label{fig2d}\includegraphics[width=0.24\textwidth]{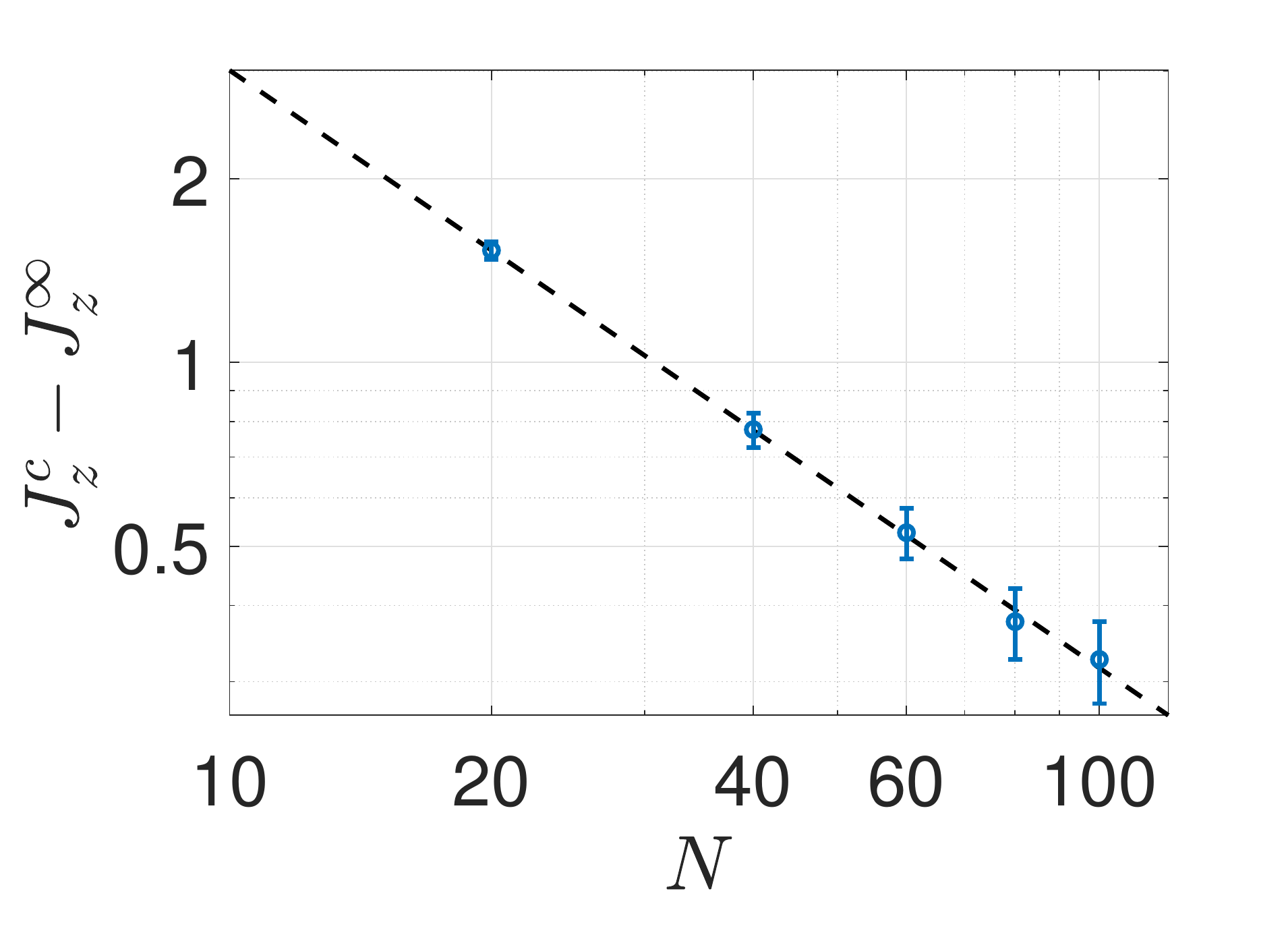}}\hfill 
\subfloat[]{\label{fig2c}\includegraphics[width=0.24\textwidth]{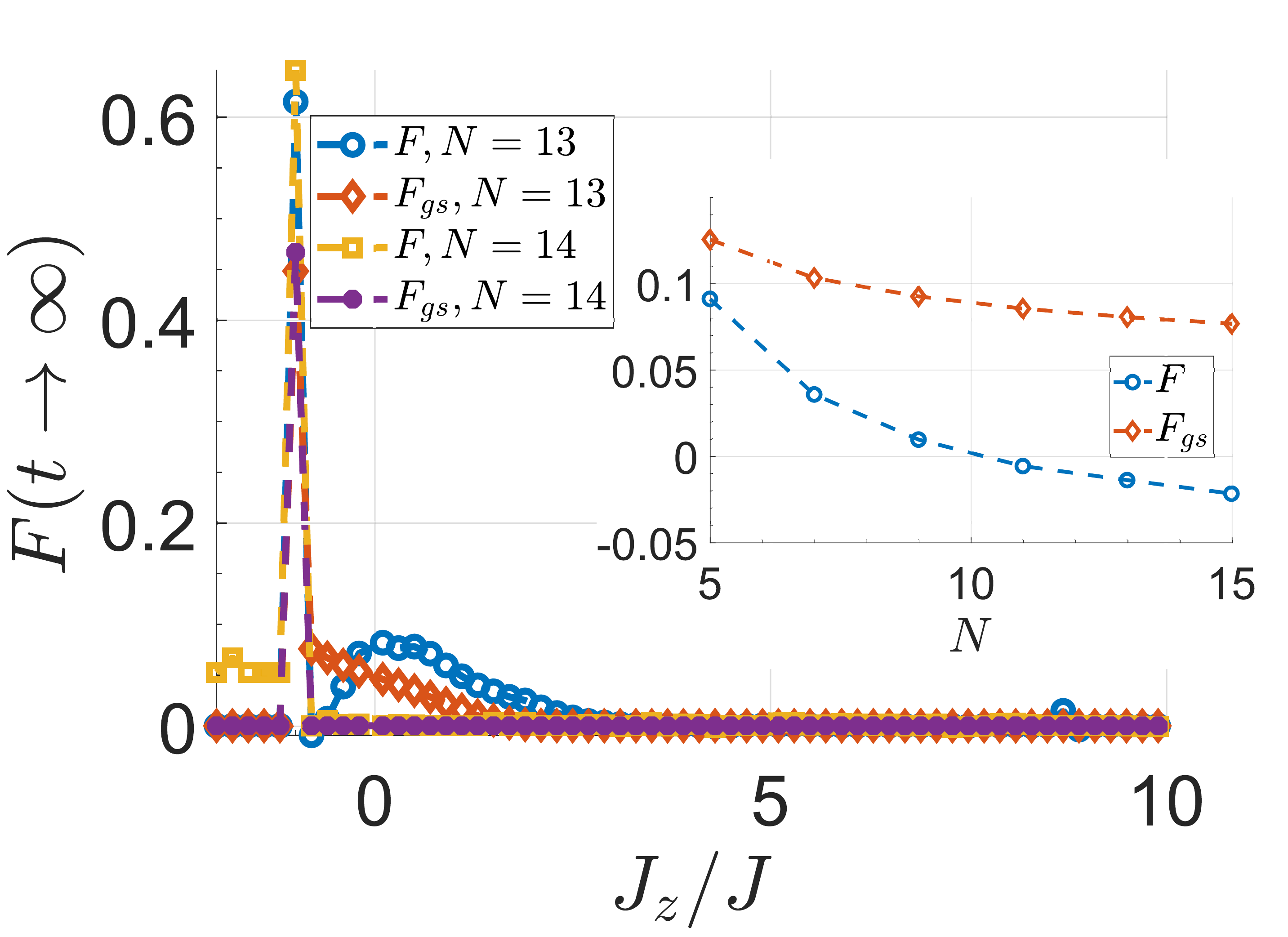}}\hfill 
\caption{(a) The OTOC saturation values for a periodic-boundary chain with $N=14$ size and a short-time of $tJ = \frac{\pi}{4}10^1$ at fields $h/J=0$ (orange-squares: Eq.~\eqref{DegsaturationEq}, purple-crosses: Eq.~\eqref{degSubs}) and $h/J=4$ (blue-circles: Eq.~\eqref{DegsaturationEq}, red-diamonds: Eq.~\eqref{degSubs}), for $\sigma^z_i$ operator. (b) Real-time dynamics (blue-circles) averaged over a time interval $tJ = 10$, $\bar{F}$, and its ground state contribution $F_{gs}$ (orange-diamonds) with DMRG algorithm and MPS for $N=60$ at $h/J=0$.  (c) System size scaling of $F_{gs}$ shows $J_z^c = a N^{\xi} + J_z^{\infty}$ with exponent $\xi=-0.98$ and $J_z^{\infty}=1.02$. (d) The OTOC saturation values for $\sigma^x_i$ operator at $h/J=0$, $N=13$ (blue-circles: Eq.~\eqref{DegsaturationEq}, red-diamonds: Eq.~\eqref{degSubs}) and $N=14$ (orange-squares: Eq.~\eqref{DegsaturationEq}, purple-crosses: Eq.~\eqref{degSubs}) for time $tJ=\frac{\pi}{4}10^3$. Inset: System-size scaling of Eq.~\eqref{DegsaturationEq} (blue-circles) and Eq.~\eqref{degSubs} (red-diamonds) at $J_z/J=-0.9$.}
\label{Fig2}
\end{figure}

We plot two cross-sections from Fig.~\ref{Fig1} in Fig.~\ref{fig2a} where the lines with orange-squares ($h/J=0$) and blue-circles ($h/J=4$) are the saturation values, Eq.~\eqref{DegsaturationEq} for a short-time $tJ\sim\frac{\pi}{4}10^1$ (long-time results in \cite{supp}). We also plot the leading order term in the saturation, $F_{\text{gs}}(t\rightarrow \infty)$ in Fig.~\ref{fig2a} with purple-cross ($h/J=0$) and red-diamond ($h/J=4$) lines. The OTOC saturation exactly reduces to the ground state contribution with no correction $F_{\text{ex}}=0$ in the Ising-ferromagnet, meaning that the saturation value in the ordered phase is exactly predicted by the Eq.~\eqref{degSubs}. The reason follows as: the spins are fully polarized in the ferromagnetic ground states, and they belong to the opposite magnetization sectors of the Hamiltonian which has magnetization conservation $[H,S_z]=0$ ($S_z=\sum_i \sigma^z_i$). Since they are the only states of their corresponding magnetization sectors, the fluctuations in the matrix elements are exactly zero, $|W_{[1,\alpha][\theta,\beta]}| = 0$. This is why the system does not scramble information at all $F(t\rightarrow \infty) = 1$, even though the XXZ model is an interacting model. We emphasize that this nonscrambling is not due to integrability of XXZ model, rather it is the signature of critical order. The rotational symmetry also protects the ferromagnetic ground states from hybridizing, all of which results in no finite-size effects at the phase boundary from ferromagnet to XY-paramagnet. In the disordered-XY phase ($h/J=0$), the ground state contribution is zero $F_{\text{gs}}=0$, leaving the correction term to dominate the saturation value, however with a small magnitude as explained above. This is the reason of the mismatch between the OTOC saturation value and its leading order term, seen in the XY-phase of Fig.~\ref{fig2a}, while we are still able to differentiate the disordered phase from the ordered phases. Finally, in the Ising-antiferromagnet the exact agreement between Eqs.~\eqref{DegsaturationEq} and \eqref{degSubs} takes place only at the $J_z/J \rightarrow \infty$ limit. As we approach the phase boundary towards the XY-phase, the fluctuations between matrix elements gradually increase, $|W_{[1,\alpha][1,\beta]}| \rightarrow 0$ \cite{supp}, result in a nonzero but small correction term to the ground-state contribution and eventually drive the phase transition. Since the finite-size effect is significant for small sizes with exact methods, we apply density-matrix renormalization group (DMRG) algorithm with matrix product states (MPS) \cite{ITensor,supp} to a system with $N=60$ and compute the real-time dynamics averaged over a short-time interval of $tJ=10$ shown with blue-circles in Fig.~\ref{fig2b} with orange-diamonds being $F_{gs}$, Eq.~\eqref{degSubs}. Note that the transition point significantly shifts towards the equilibrium phase transition point, $J_z/J=1$. We extract the system-size scaling parameters from our DMRG computations, Fig.~\ref{fig2d} and observe that the system indeed approaches to the equilibrium transition point when $N\rightarrow \infty$, $J_z^c = a N^{\xi} + J_z^{\infty}$ with exponent $\xi=-0.98$ and $J_z^{\infty}=1.02$ with a power-law scaling.

We plot the OTOC with $\sigma^x_i$ operator for $N=13$ (blue-circles) in Fig.~\ref{fig2c}: the OTOC saturation remains small in all three phases and thus the OTOC can hardly distinguish the XY-ordered from XY-disordered phases. When the chains with even number of spins are used ($N=14$, orange-squares) in the theory, we do not even obtain any difference between the phases. This is in agreement with our theoretical predictions discussed previously. Additionally, the fluctuations between the matrix elements of quasi-long range order parameter $\sigma^x_i$ are always maximal regardless of the phase. Hence, we observe the mismatch between the OTOC saturation and its ground state contribution (red-diamonds $N=13$ and purple crosses $N=14$). The inset of Fig.~\ref{fig2c} shows that the OTOC saturation value and its ground state contribution both decrease with the system size for odd-numbered chains, exhibiting that the OTOC saturation cannot capture the quasi-long range order in bigger systems and thermodynamic limit. We briefly note that the detection of the order at $J_z/J=-1$ is robust due to the massive degeneracy in the ground state at this point of different symmetry (SU(2) symmetry). 

\emph{Conclusion.} Our theoretical predictions on the XXZ model can be experimented with cold atoms \cite{PhysRevLett.91.090402}. Based on the studies in the literature \cite{PhysRevLett.121.016801,2018arXiv181201920W,2018arXiv181111191S} and our results in the XXZ model, our method seems to be universal in explaining the reasoning behind the relation between scrambling and the quantum criticality. In this sense, our method is an analogue of the Eigenstate Thermalization Hypothesis: It tells us the criteria of how scrambling probes criticality; though it is independent of the integrability of the system, unlike ETH. Dynamical decomposition of OTOC is a complementary tool to the real-time evolution of a state in determining the OTOC saturation value. However in addition to providing the saturation value, it also presents us the conditions for OTOC to show either order or disorder. Based on this fact, the leading order term in our theory, Eq.~\eqref{degSubs}, could mark the phase transition points via system-size scalings. In conclusion, given that the initial state of OTOC is a state where the phase transition is expected to happen and the off-diagonal matrix elements of the operator are sufficiently suppressed in this state (or degenerate state subspace), OTOC could be used to dynamically detect the quantum phases with long-range order and capture the symmetry-breaking quantum phase transitions.
\begin{acknowledgements}
\emph{Acknowledgements.} This work was supported by National Science Foundation under Grant EFRI-1741618, the AFOSR Multidisciplinary University Research Initiative program and the ARL CDQI program. C.B.D. thanks P. Myles Eugenio for interesting discussions and comments on the manuscript.
\end{acknowledgements}

\bibliographystyle{apsrev4-1}
%\bibliography{Bibliography} % The references (bibliography) information are stored in the file named 


\pagebreak

%\widetext
\begin{center}
\textbf{\large Supplementary: Detection of quantum phases via out-of-time-order correlators}
\end{center}
%%%%%%%%%% Merge with supplemental materials %%%%%%%%%%
%%%%%%%%%% Prefix a "S" to all equations, figures, tables and reset the counter %%%%%%%%%%
\setcounter{equation}{0}
\setcounter{figure}{0}
\setcounter{table}{0}
\setcounter{page}{1}
\makeatletter
\renewcommand{\theequation}{S\arabic{equation}}
\renewcommand{\thefigure}{S\arabic{figure}}
\renewcommand{\bibnumfmt}[1]{[S#1]}
\renewcommand{\citenumfont}[1]{S#1}
%%%%%%%%%% Prefix a "S" to all equations, figures, tables and reset the counter %%%%%%%%%%

\section{Energy-time relation and finite-size effects}\label{sec1} 

Eq.~5 in the main text clearly exhibits how the OTOC can suffer from the finite-size effects. Imagine that the finite-size lifts the degeneracy. Based on the dynamical equation of OTOC,
\begin{eqnarray}
F(t) = \sum_{\alpha,\beta,\gamma,\gamma'} c_{\alpha}^* b_{\beta} e^{-i (E_{\beta}-E_{\alpha} + E_{\gamma}-E_{\gamma'})t} W^{\dagger}_{\alpha \gamma} V^{\dagger}_{\gamma \gamma'} W_{\gamma' \beta},\notag
%\label{OTOCdynamics}
\end{eqnarray}
one would write the ground state contribution to the OTOC at zero temperature as
\begin{eqnarray}
\label{degeneracyLifting}
F_{gs}(t)&& \sim |W_{[1,1][1,2]}W_{[1,2][1,1]}|^2\text{exp}\left[-2i(E_{[1,2]}-E_{[1,1]})t\right].\notag \\
\end{eqnarray}
Since Eq.~\eqref{degeneracyLifting} is the dominant contribution to the OTOC in the ordered phase, the order will eventually be invisible to the saturation value, and it will be encoded in the frequency spectrum of the OTOC. However since this is a finite-size effect, we expect to see finite saturation value for all times in the thermodynamic limit. The period of the emerging oscillation (due to degeneracy-lifting) is $\tau=\pi/(E_{[1,2]}-E_{[1,1]})$. Then starting from $t\sim \tau/2$, the order will be invisible to the saturation value. Thus, the region where the system seems to have reached its most correlated state before the finite-size effects show up, could be defined for $t\ll \tau/2$; whereas the order will be most visible to the frequency spectrum around $t \gg \tau/2$. Further, the ground state contribution will exist in the saturation value as a non-zero effect for a time $t \sim \tau/4$, where the time-averaging will reveal the order. The relation between evolution time and energy spectrum reflects the observation that longer the time evolution, better the resolution of the energy spectrum. This, in turn, helps us to estimate the time interval of the corresponding dynamics simulation of the theory Eq.~5 (in the main text), even though Eq.~5 is explicitly time-independent. 

\begin{figure}
\centerline{\includegraphics[width=0.45\textwidth]{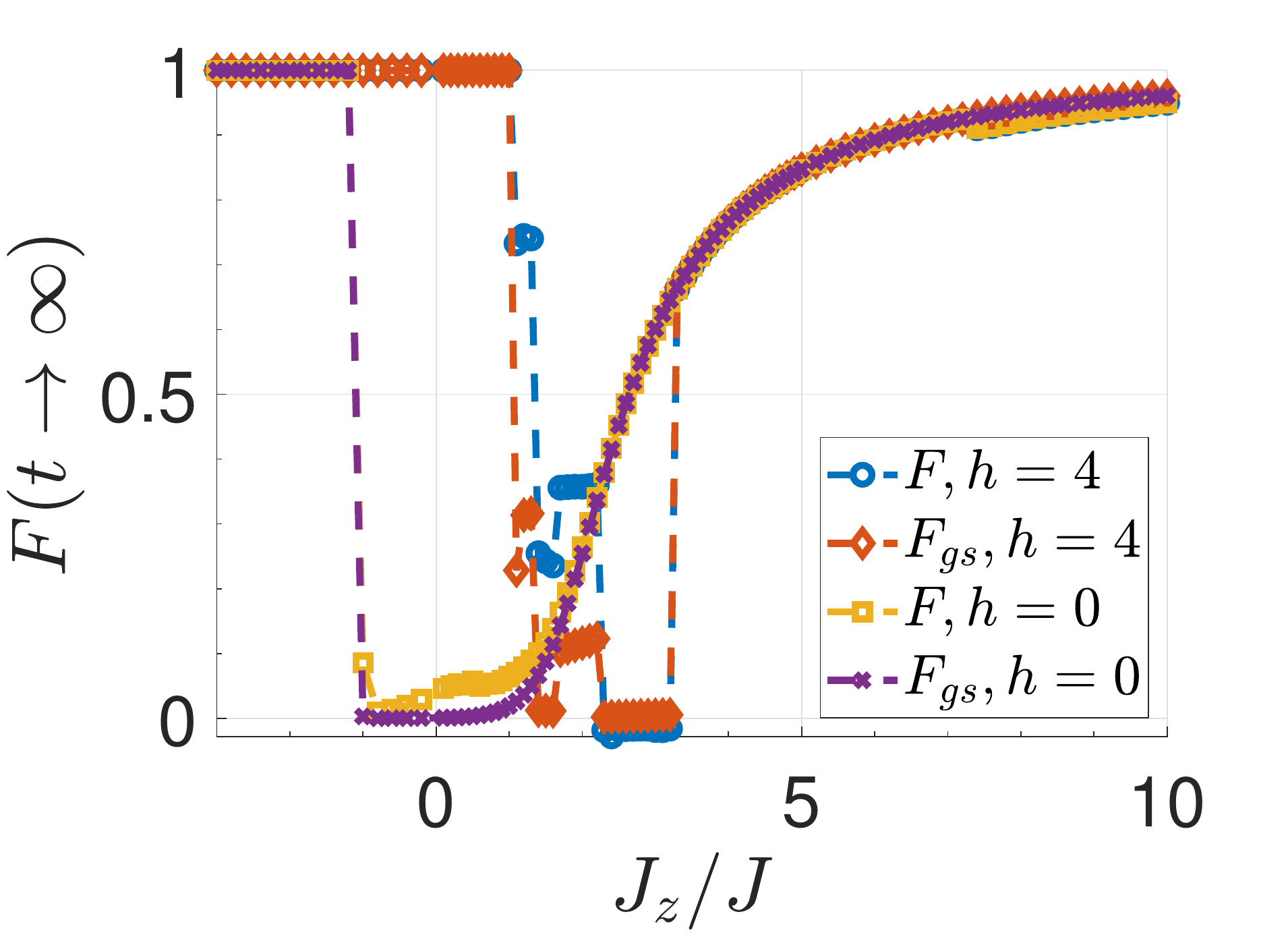}}
\caption{The OTOC saturation values for an open-boundary chain with $N=13$ size and a long-time of $tJ \sim \frac{\pi}{4}10^3$ at fields $h/J=0$ (orange-squares: Eq.~4, purple-crosses: Eq.~6 and $h/J=4$ (blue-circles: Eq.~4, red-diamonds: Eq.~6, for $\sigma^z_i$ operator.}
\label{SuppFig4b}
\end{figure}

\begin{figure}
\centerline{\includegraphics[width=0.45\textwidth]{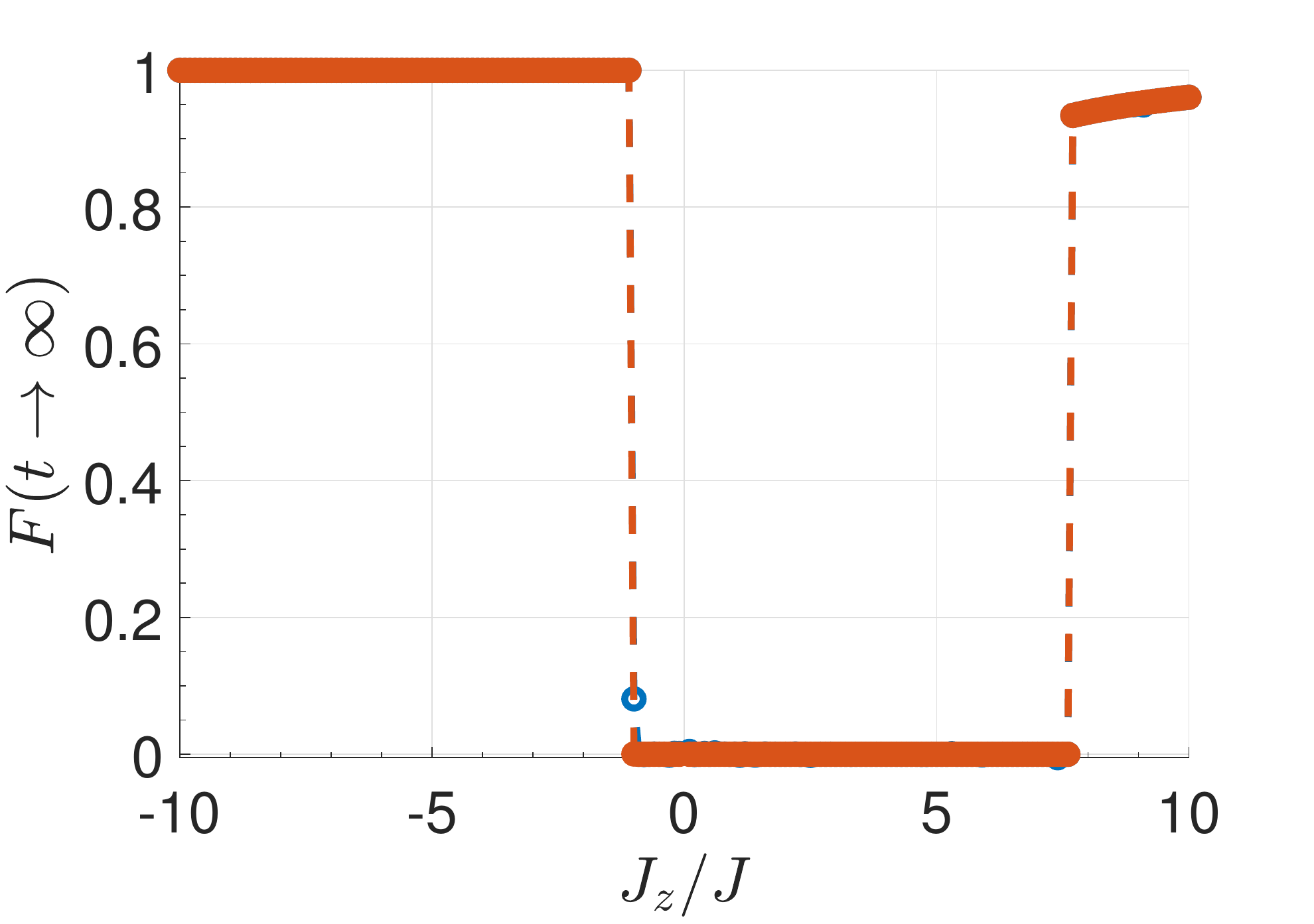}}
\caption{OTOC saturation value (Eq.~(3), blue line) and its ground state contribution (Eq.~(5), orange line) for $h/J=0$, $N=14$ and for time $tJ \lesssim\frac{\pi}{4}10^3$. The anti-ferromagnetic order is concealed due to finite-size effects appearing in long-times.}
\label{SuppFig4}
\end{figure}

For a phase transition that involves antiferromagnetic order, due to the doubling of the unit cell size, it is not uncommon that the finite-size contributions may oscillate as we increase the system size, depending on whether the system is composed of odd (Fig.~\ref{SuppFig4b}) or even (Fig.~2a in the main text) number of sites. For periodic boundary conditions, the systems with even number of sites usually show smaller finite-size effects. In our studies, though, we observe the opposite: chains with odd number of spins experience finite-size effects less. As far as the OTOC is concerned, stronger finite size effects are expected for systems with even number of sites, since the key to obtain the OTOC saturation value is to sum over all the quantum states in the ground state subspace. In an Ising-ordered phase, an exact two-fold degeneracy is guaranteed for a chain with odd number of spins by the Kramers degeneracy theorem, because an odd number of spin-1/2 results in a half-integer total spin. For a system with even number of sites (i.e. integer total spin), such an exact degeneracy is not expected and thus the degeneracy is lifted by finite size effects more strongly than chains with odd number of spins. This explains the dramatic difference between the results of open and periodic boundary conditions at the XY-antiferromagnet boundary. We also note that the ground states belong to the $S_z=0$ magnetization sector in the antiferromagnet, which is the biggest sector of the Hamiltonian and hence they would hybridize with each other. To alleviate the finite-size effects, we make use of the time-energy relation explained above and plot the OTOC for significantly smaller interval of time, $tJ\sim\frac{\pi}{4}10$ in the main text. This should be compared with the results of a long-time evolution $tJ \sim \frac{\pi}{4}10^3$ in Fig.~\ref{SuppFig4}. This comparison is a good example of how finite-size effects could show up in scrambling, restricting the order to short-times. We also note that even though our method is valid at the infinite-time limit, due to the energy-time relation employed in the computations it produces sufficiently good results for the real-time dynamics in a short-time evolution.

\begin{figure}
\centerline{\includegraphics[width=0.45\textwidth]{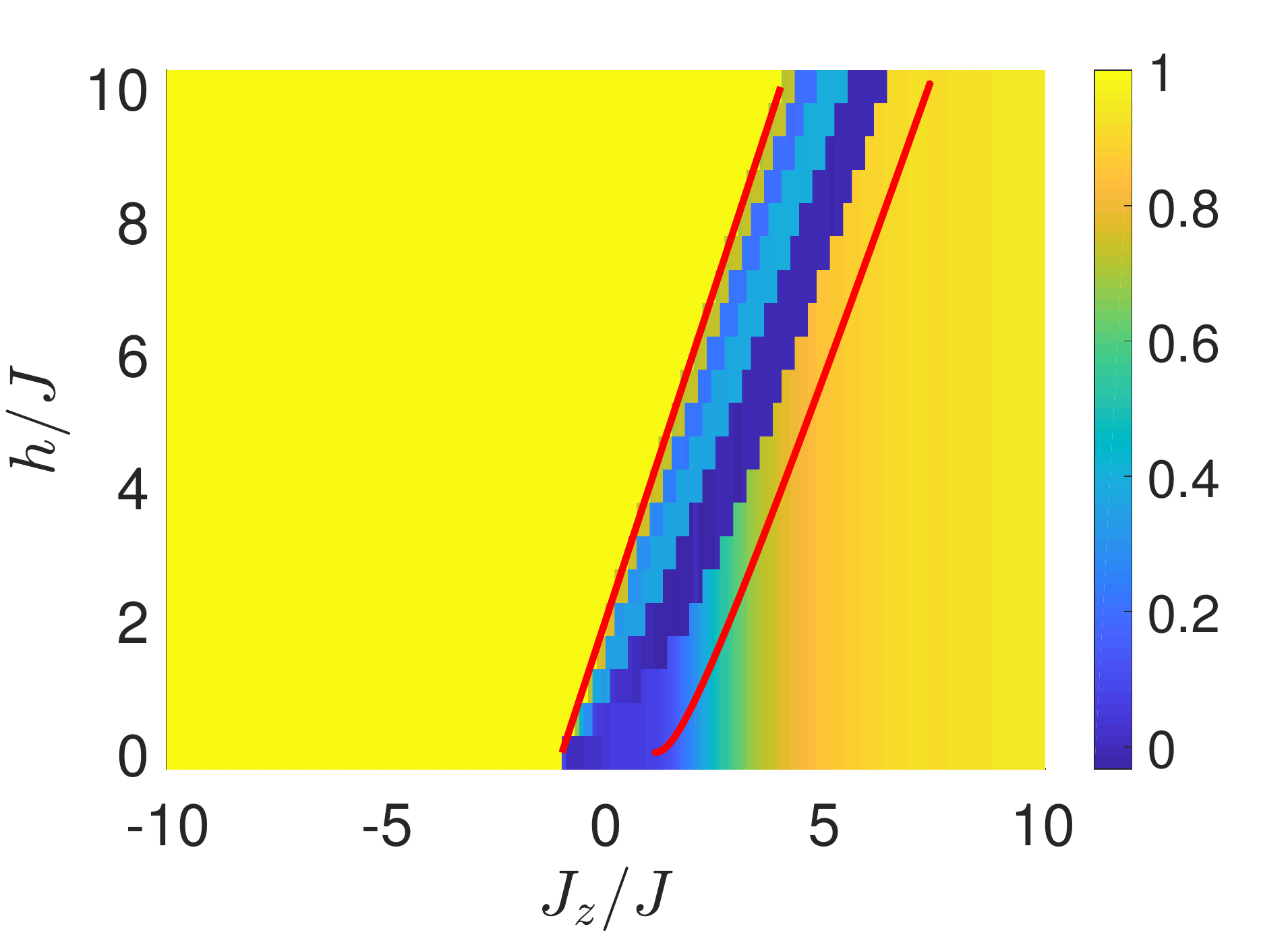}}
\caption{The OTOC phase diagram (via Eq.~(3)) for odd-numbered chains while the x-axis is the spin interaction strength in the z-direction $J_z/J$ and y-axis is the magnetic field $h/J$, for $N=13$ system size and $\sigma_z^i$ where the observation spin is chosen from bulk, when open boundary conditions are set and initial state is a ground state. The time-scale where the results are valid is $tJ \sim \frac{\pi}{4}10^3$.}
\label{SuppFig5}
\end{figure}
\begin{figure}[tbp]
\centering
\subfloat[]{\label{fig:fig1a}\includegraphics[width=0.24\textwidth]{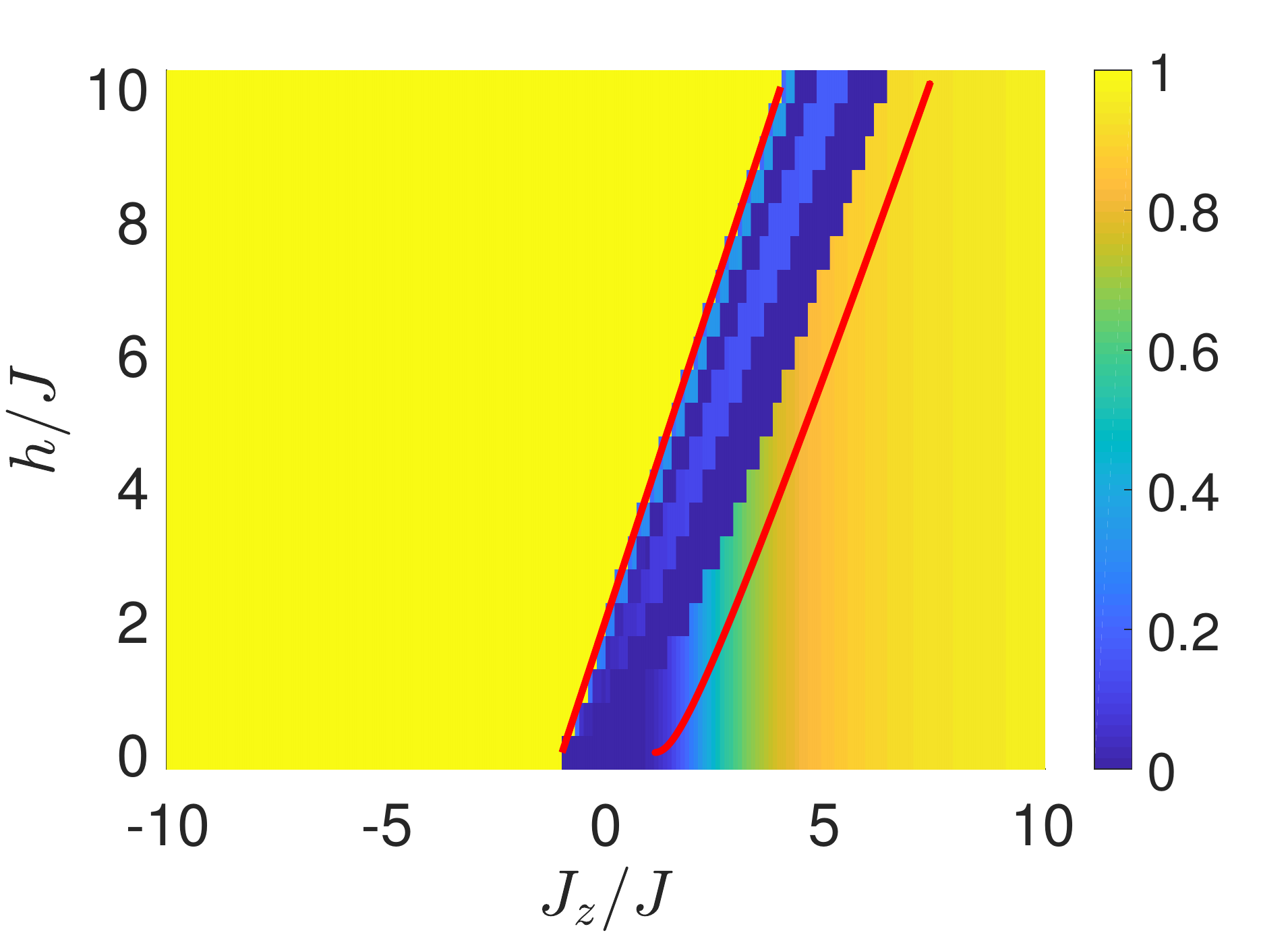}}\hfill \subfloat[]{\label{fig:fig1b}%
\includegraphics[width=0.24\textwidth]{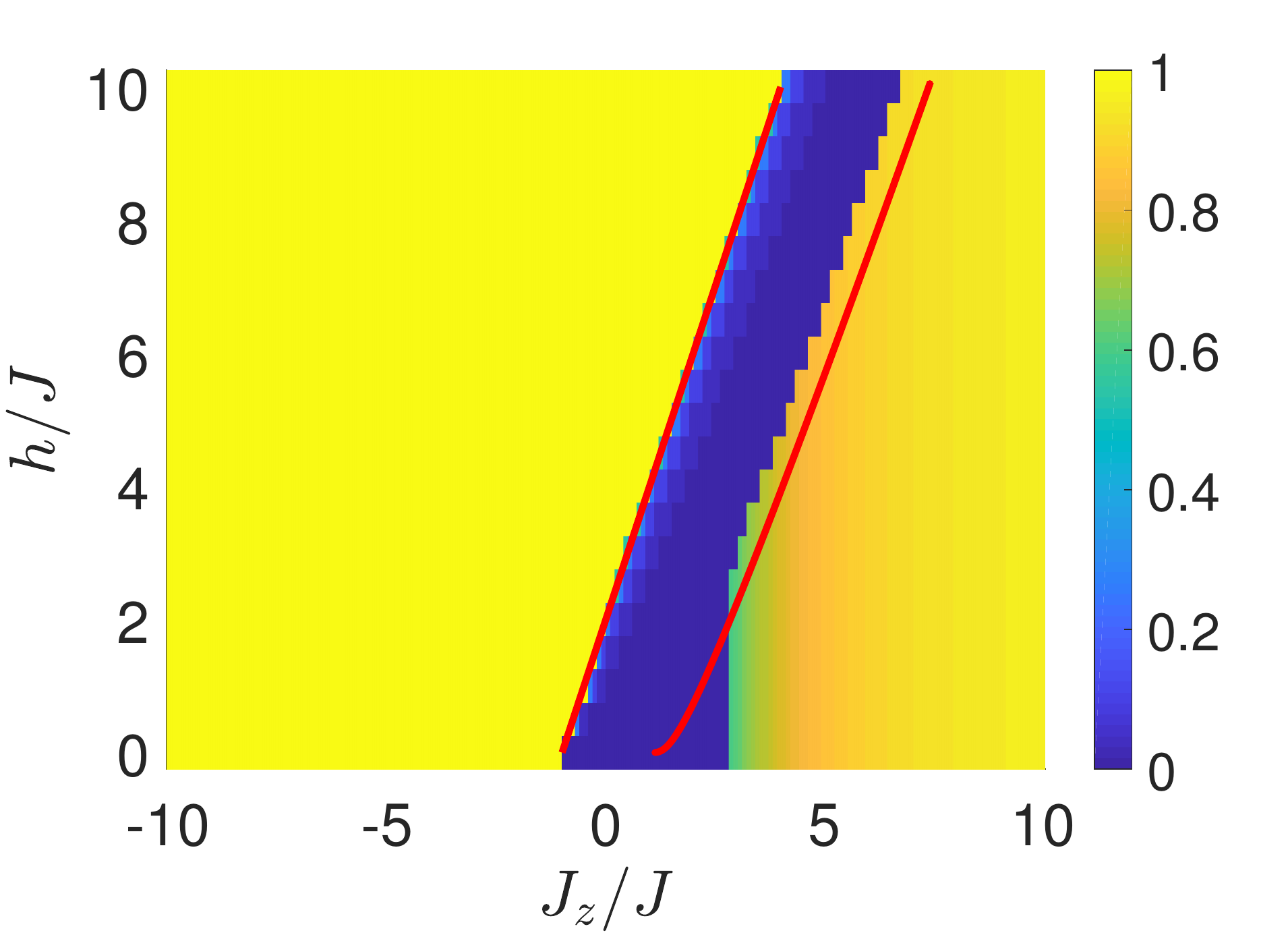}}
\hfill 
\caption{Ground state value contribution (Eq.~(5)) to OTOC for (a) odd-numbered $N=13$ and (b) even-numbered $N=14$ chains with respect to $J_z/J$ at x-axis and field $h/J$ at y-axis.}
\label{SuppFig6}
\end{figure}
Fig. \ref{SuppFig5} is the phase diagram of odd-numbered chains with phase boundaries dictated by the Bethe ansatz for an infinite-size chain. Even though the finite-size effects for small fields are more severe in even-numbered chains than odd-numbered chains in small systems, we note that the transition in odd-numbered chains is ambiguous. The continuous transition between XY- and antiferromagnetic phases due to the nature of the doubly-degenerate ground states in odd-numbered chains (as explained above), prevents a straightforward system-size scaling based on odd-numbered chains. Therefore we chose to focus on even-numbered chains in the main text. We also note that as we compute results for bigger system sizes, the apparent odd-even effect disappears. The difference between Bethe ansatz results and the OTOC phase boundary for the anti-ferromagnetic to XY phase in high fields is also due to finite-size effects. This could be seen in Figs. \ref{SuppFig6}, where we compare the ground state value in OTOC with the exact phase boundaries. This means that the ground state contribution similarly suffers from the finite-size effects as well. Hence the results point to the effect of finite-size on the ground state manifold rather than the incapability of OTOC to probe the phase transition as precisely as exact results. It is an interesting question how the system size scaling results of OTOC and its ground state contribution would compare with the existing methods of determining the phase boundary, e.g. Binder ratio, fidelity measures, determining the energy gap, spatial correlation functions etc.

\section{Operator Ansatz Demonstrated}

\begin{figure}
\centerline{\includegraphics[width=0.45\textwidth]{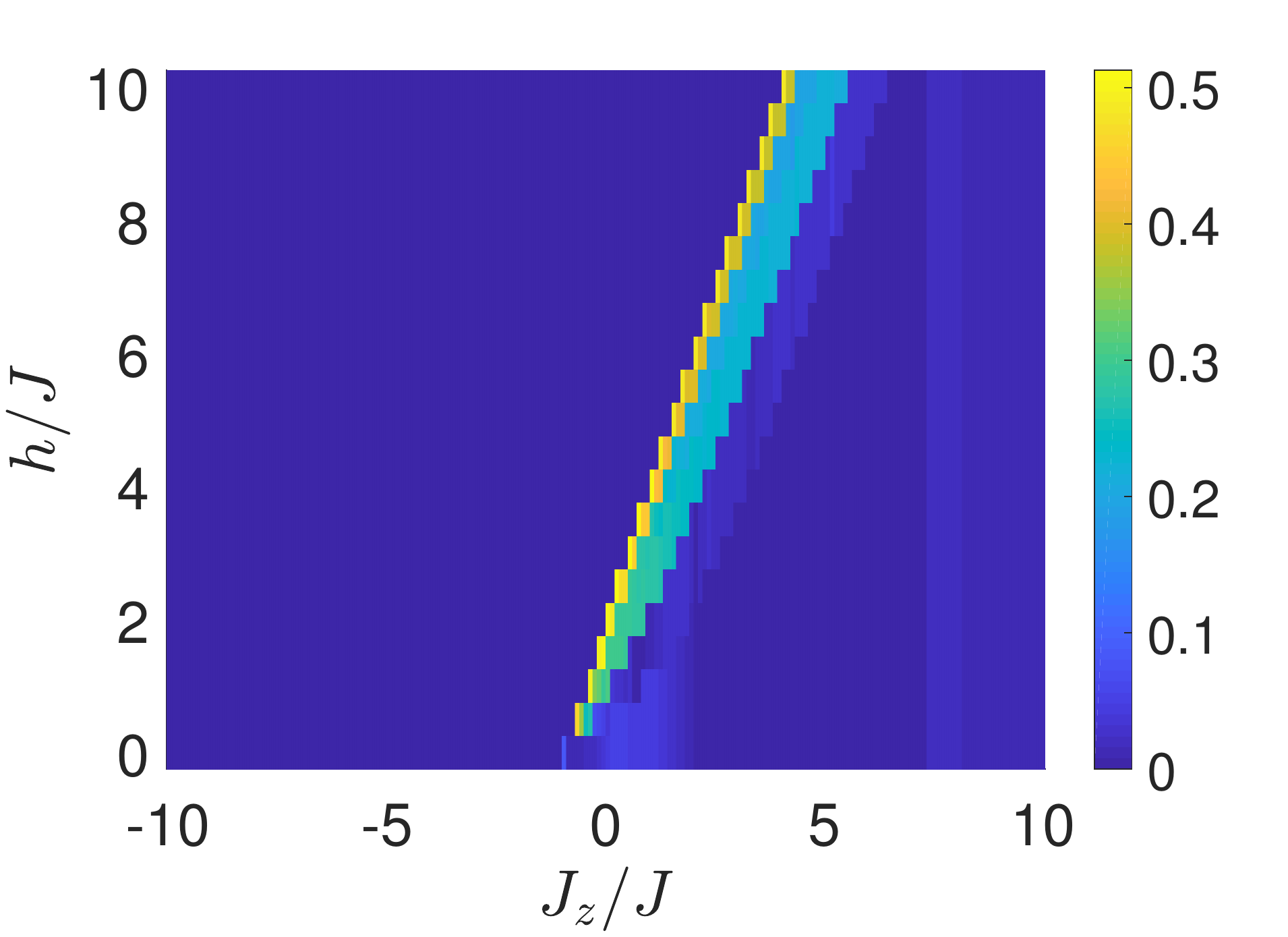}}
\caption{The difference between the OTOC saturation values (via Eq.~(3) in main text) and the ground state contribution for the phase diagram while the x-axis is the spin interaction strength in the z-direction $J_z$ and y-axis is the magnetic field $h$, for $N=13$ system size and $\sigma_z^i$ where the observation spin is chosen from bulk, when open boundary conditions are set and the initial state is a ground state.}
\label{SuppFig1}
\end{figure}

Here we give the additional results of XXZ model on the relation between OTOCs and phase transitions. Fig. \ref{SuppFig1} shows the difference between the OTOC saturation values, Fig.~\ref{SuppFig5} and the ground state contribution in these values, Fig.~\ref{fig:fig1a}. The mismatch between OTOC saturation value and ground state contribution to it is clear in XY-phase, due to the fact that the correction term of the excitations is dominant in the XY-phase, as explained in the text.

\begin{figure}
\centering
\subfloat[]{\label{fig:fig4a}\includegraphics[width=0.24\textwidth]{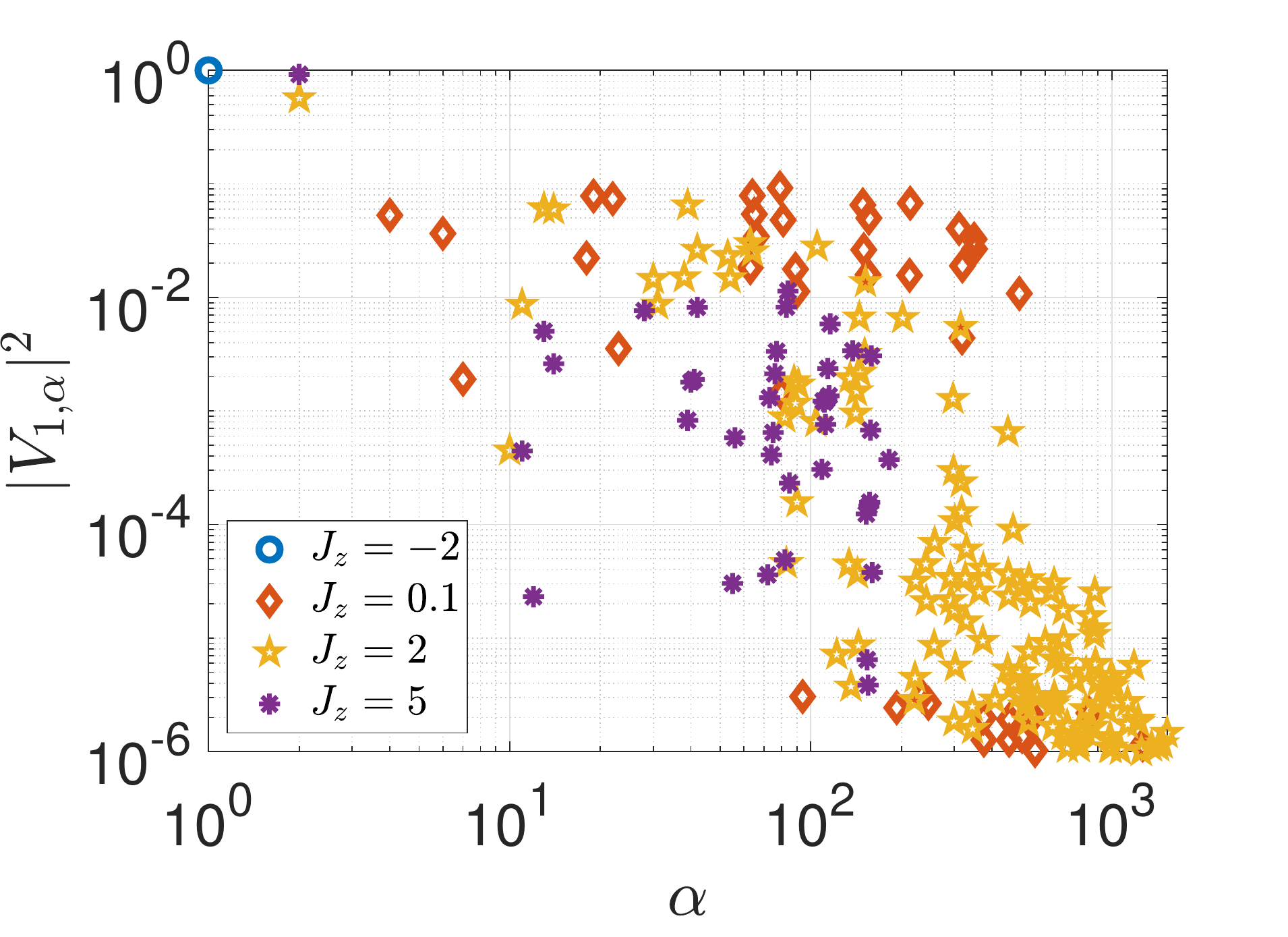}}\hfill \subfloat[]{\label{fig:fig4b}\includegraphics[width=0.24\textwidth]{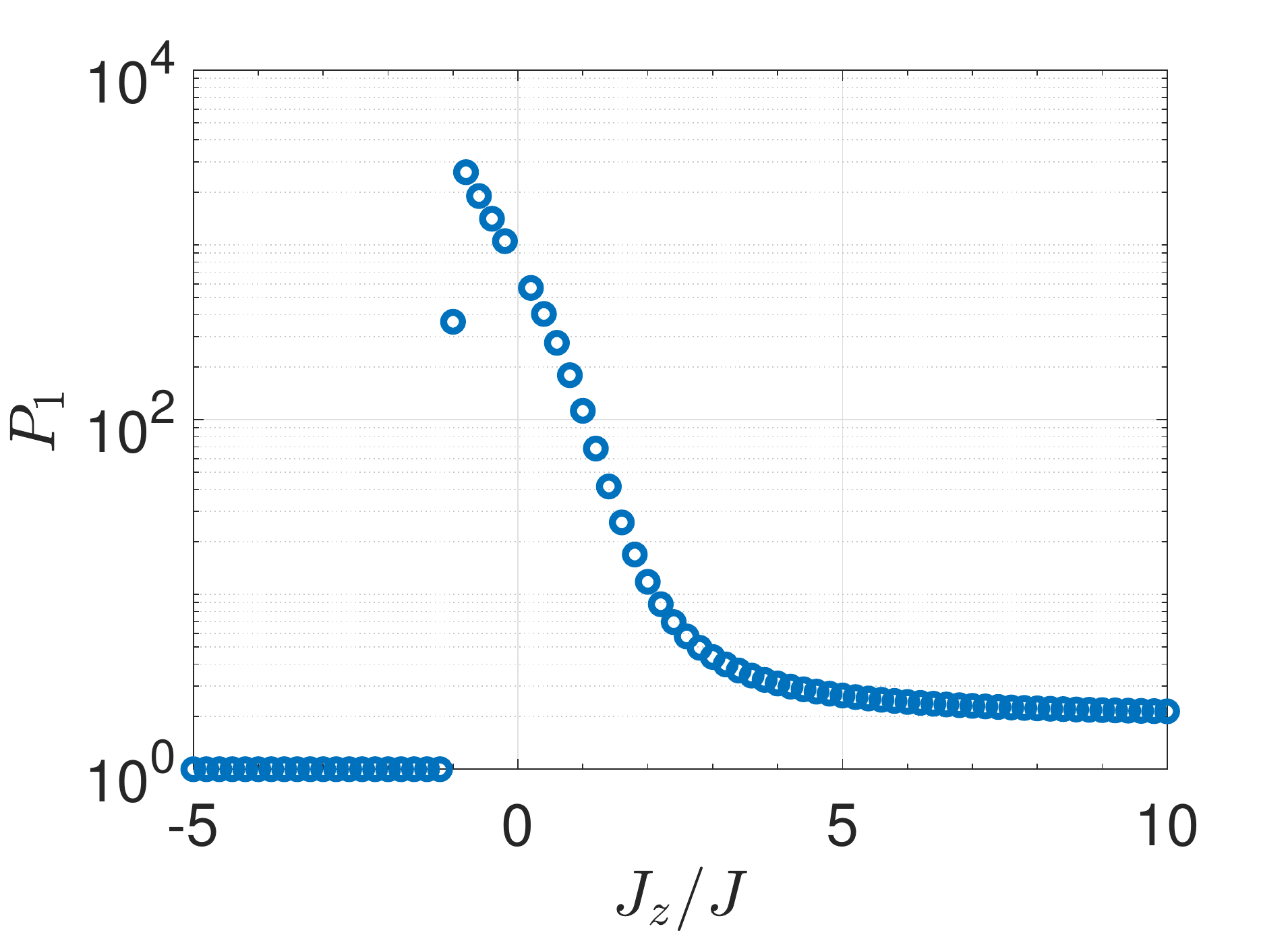}}\hfill 
\subfloat[]{\label{fig:fig4c}\includegraphics[width=0.24\textwidth]{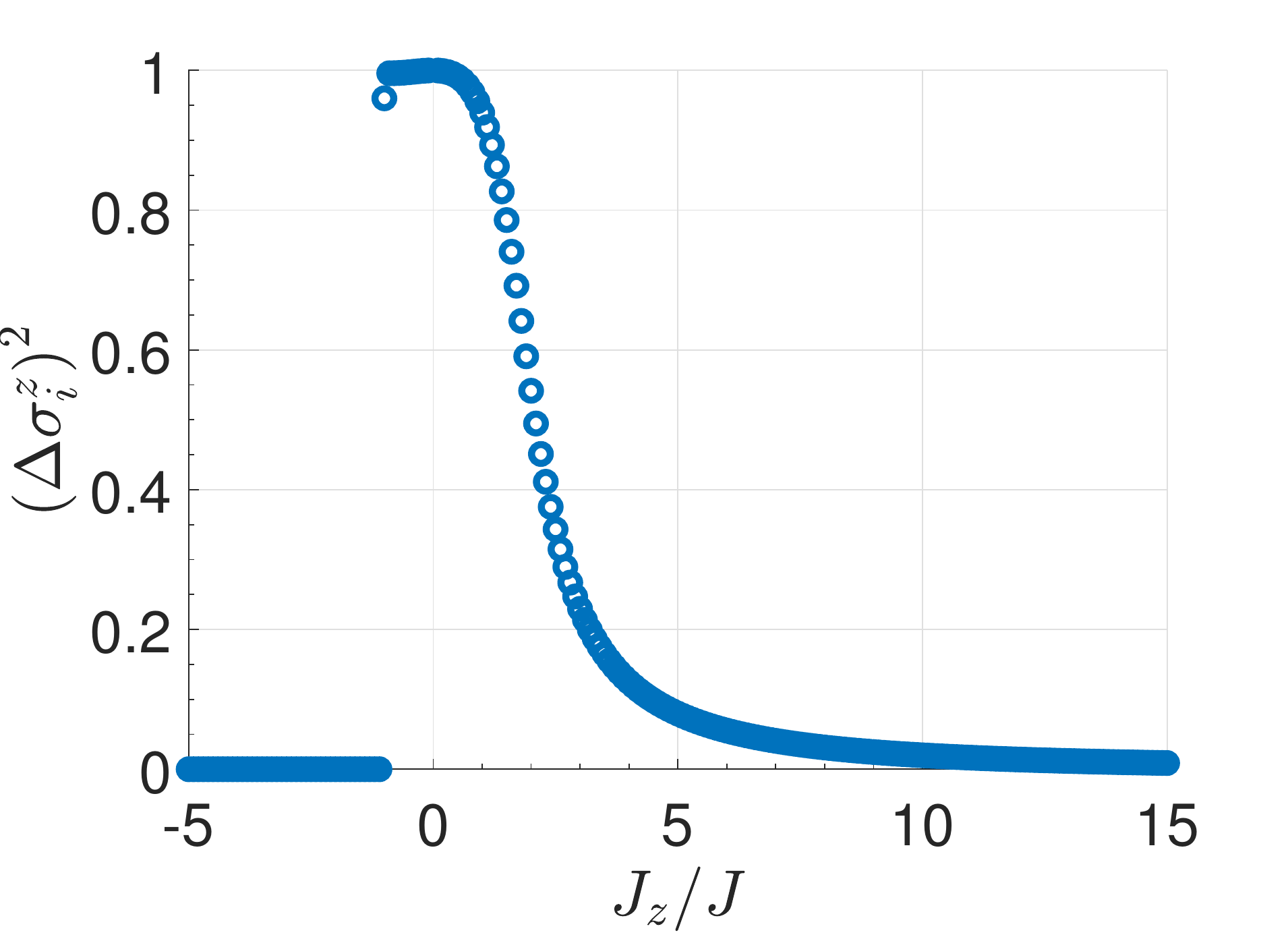}}\hfill \subfloat[]{\label{fig:fig4d}\includegraphics[width=0.24\textwidth]{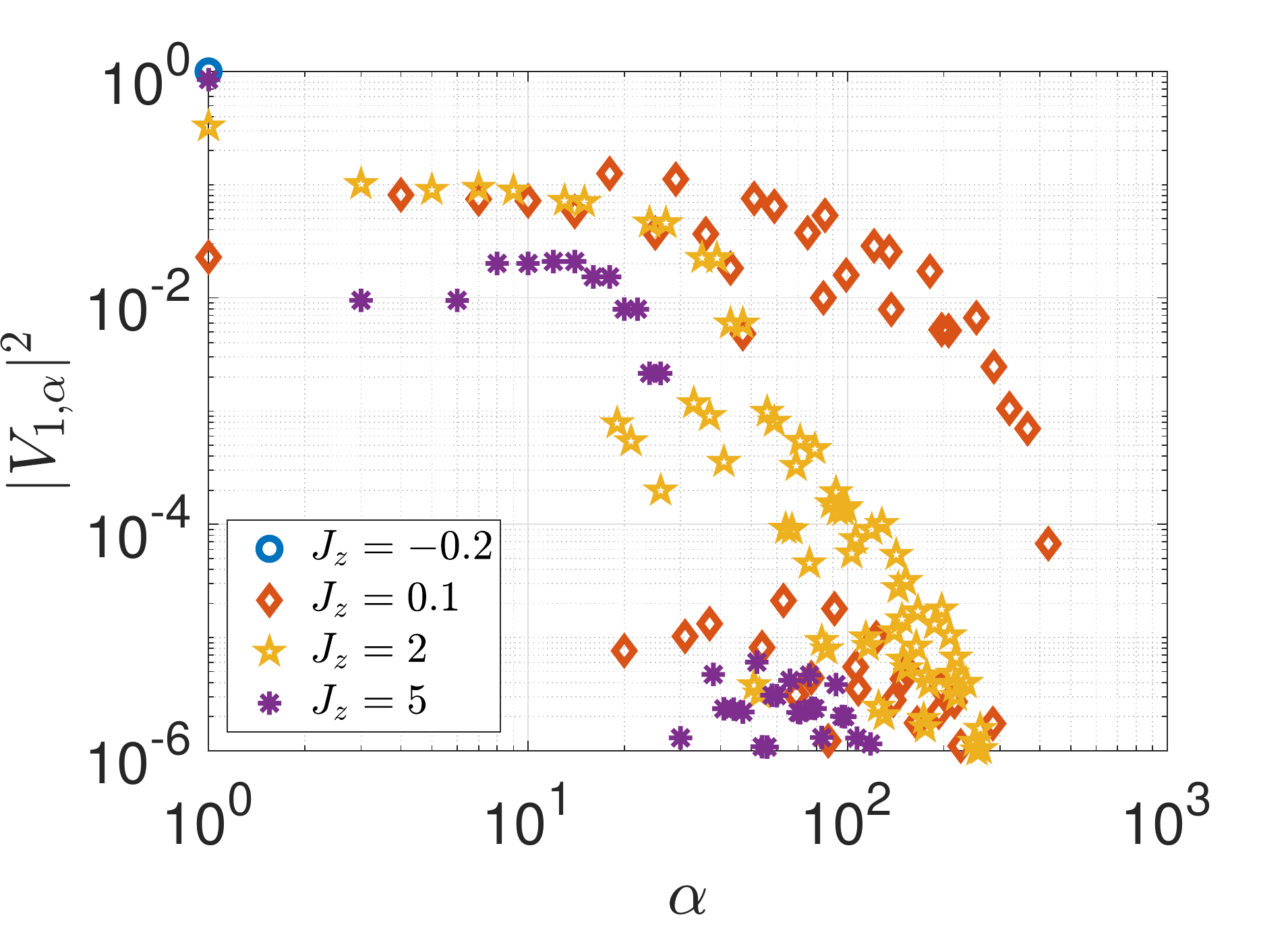}}\hfill 
\caption{(a) Matrix elements of operator $\sigma_z^i$ of a bulk spin for a ground state $|V_{1,\alpha}|^2$ at $N=14$ for various $J_z$: $J_z/J=-2$ (blue-circles), $J_z/J=0.1$ (red-diamonds), $J_z/J=2$ (yellow-pentagons) and $J_z/J=5$ (purple-dots). (b) Participation ratio of the ground state with respect to $J_z/J$ for a system size $N=14$, while the reference basis is spin basis. (c) Fluctuations of a ground state at $N=15$. (d) The matrix elements of operator $\sigma_z^i$ of a bulk spin for a ground state $|V_{1,\alpha}|^2$ at $N=13$ (odd-numbered chains) for various $J_z$: $J_z/J=-2$ (blue-circles), $J_z/J=0.1$ (red-diamonds), $J_z/J=2$ (yellow-pentagons) and $J_z/J=5$ (purple-dots).}
\label{SuppFig3}
\end{figure}

Fig. \ref{fig:fig4a} shows the matrix elements of the long-range order bulk operator $\sigma_z^i$ used in the study only for a ground state $|V_{1,\alpha}|^2$. Note how the operator's matrix elements satisfy the operator ansatz put forward in the text: in Ising-ordered phases we observe $|W_{[1,\alpha][1,\beta]}|^2 \gg |W_{[1,\gamma][\theta,\gamma']}|^2$ while in the Ising-disordered phase (XY-phase) $W_{[1,\alpha][1,\beta]} \sim 0$ and $|W_{[1,\alpha][\theta,\beta]}|^2 \ll 1$ is satisfied. Fig. \ref{fig:fig4b} shows the participation ratio (PR) value of the ground state in terms of spin basis. PR is defined as
\begin{eqnarray}
P_{\alpha} &=& \left( \sum_{n=1} |\psi_{\alpha n}|^4 \right)^{-1},
\end{eqnarray}
where $\alpha$ are eigenstates and $n$ are the reference basis. PR is a measure of fluctuations of a state in a reference basis. We see that the ferromagnetic ground states ($J_z/J < -1$) are more localized compared to anti-ferromagnetic ground states ($J_z/J > 1$), because of the subspaces that they belong to under a $S_z$ conserving Hamiltonian. As a result, anti-ferromagnetic ground states are more susceptible to both finite-size effects (mixing in energy levels) and the effect of the rest of the terms in the Hamiltonian. This is also the reason why OTOCs are better in capturing the transition from a ferromagnet to a XY-paramagnet compared to anti-ferromagnet to XY-paramagnet. Unless $J_z \gg J$, the XX- and YY-coupling terms cause the Neel states to slightly couple to the other states in $S_z=0$ subspace. 

The operator ansatz on the matrix elements can also be seen in terms of the fluctuations in a ground state. The fluctuations of the ground state can be defined as $\left(\Delta \sigma_z^i\right)^2 =  \Braket{(\sigma_z^i)^2} - \Braket{\sigma_z^i}^2$ where the expectation is taken over the ground state and seen in Fig.~\ref{fig:fig4c} for a system size of $N=15$. Hence, we state that the fluctuations are maximized in XY-phase, causing a dominant correction term in the OTOC saturation value. The fluctuations are zero in the ferromagnetic phase and they approach to zero in the anti-ferromagnetic phase as $J_z/J \rightarrow \infty$. Note that this is the case because an open boundary chain with odd-number of spins have two ground states with different $S_z$ quantum numbers in the anti-ferromagnetic phase. This can be seen more explicitly in the matrix elements of an odd-numbered chain in Fig.~\ref{fig:fig4d}. The operator ansatze are satisfied as expected, however note that the main contribution comes from the diagonal elements in the Ising-ordered phases, unlike in the even-numbered chains in Fig.~\ref{fig:fig4a}. Finally we note that the fluctuations are always maximum for the quasi-long range operator $\sigma_x^i$.

\section{Comparison of real-time dynamics with theory prediction in short times}

\begin{figure}
\centering
\subfloat[]{\label{supfig:fig8a}\includegraphics[width=0.24\textwidth]{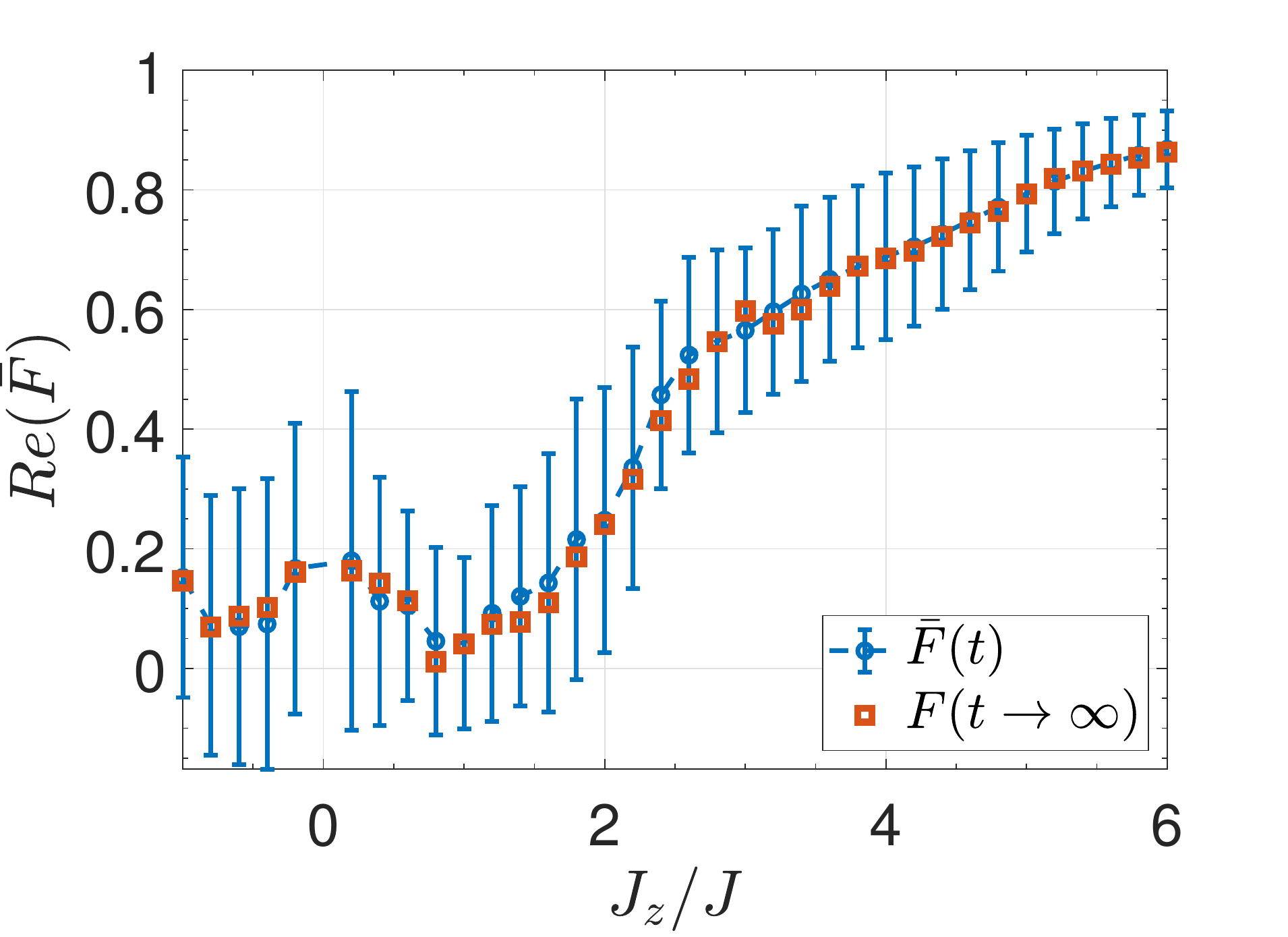}}\hfill \subfloat[]{\label{supfig:fig8b}%
\includegraphics[width=0.24\textwidth]{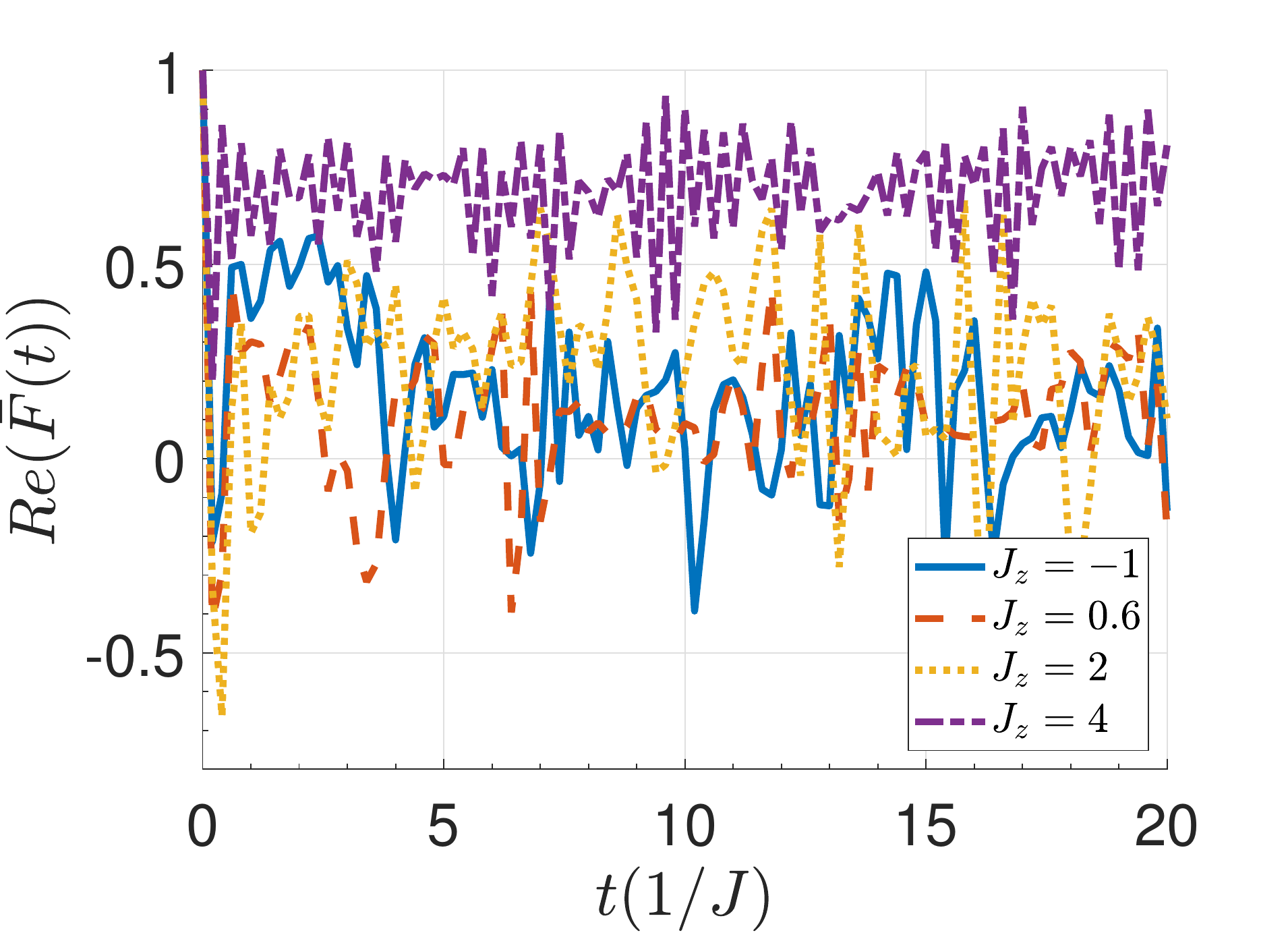}}
\hfill 
\caption{(a) The time-average of OTOC signal over a time interval of $tJ=20$ (blue circles) and the extend of the oscillations around the average with the error bars; the theory saturation value prediction Eq.~3 in the main text (red squares). (b) The real-time dynamics shown for $J_z=-1$ (blue-solid), $J_z=0.6$ (red-dashed), $J_z=2$ (yellow-dotted) and $J_z=4$ (purple-dashed dotted). Both subfigures are for a system size of $N=13$ and at a zero field $h/J=0$}.
\label{SuppFig8}
\end{figure}

Here we share a direct comparison between real-time dynamics of OTOC in short-time and the infinite-time saturation value that is predicted by Eq.~3 in the main text, to demonstrate that the analytical framework to predict the saturation value (or the time-average) is robust in finite times. Fig.~\ref{supfig:fig8a} shows the average of the time-signal (over a time interval of $tJ=20$ ) with blue circles and the extend of oscillations with the error bars around the blue circles. Some of these real-time OTOC signals can be seen in Fig.~\ref{supfig:fig8b}. Even though these signals are oscillatory and show transient features, our theory could predict the average of the signals with a good accuracy as seen with red squares in Fig.~\ref{supfig:fig8a}. Therefore, our theory is not restricted to long times. The main reason of this robustness is due to the energy-time relation we employ in the computations, hence the saturation value Eq.~3 (in the main text) changes with the interval of time-averaging even though the Eq.~3 is explicitly time-independent (see supplement Sec.~\ref{sec1}). 

\section{OTOC with odd number of spins and periodic boundary conditions}

\begin{figure}
\centerline{\includegraphics[width=0.45\textwidth]{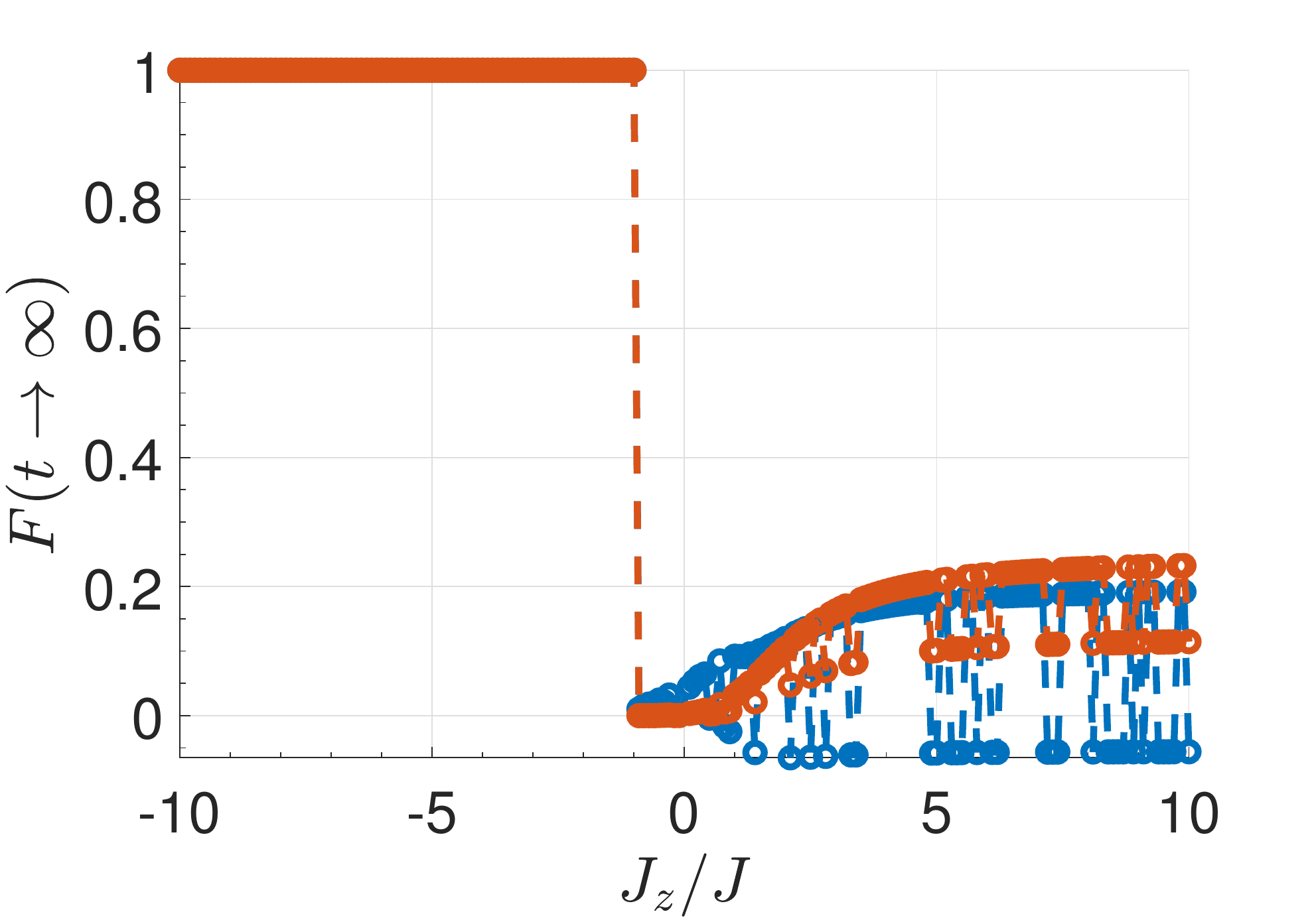}}
\caption{OTOC saturation value (Eq.~(3), blue line) and its ground state contribution (Eq.~(5), red-circles), for $h/J=0$ when $N=13$ is set and for time $tJ\lesssim\frac{\pi}{4}10^3$ with periodic boundary conditions.}
\label{SuppFig7}
\end{figure}
We present the result for odd-numbered chain if periodic boundary condition is applied, Fig. \ref{SuppFig7}. The low values and fluctuations in the anti-ferromagnetic region are a sign of how OTOC is sensitive to emerging domain walls in the ground state. Domain walls are expected in the anti-ferromagnetic ground state because of the frustration in an odd-numbered periodic chain.

\section{The remarks on the saturation value of OTOC}

Eq.~2 in main text can show why quantum chaotic spin systems should eventually decay to zero when ETH is evoked up to some approximations. When a system follows ETH, there are two criteria to satisfy: (i) $ V_{\gamma \gamma'} \ll V_{\gamma \gamma} $, where $\gamma \neq \gamma'$, and (ii) $V_{\gamma \gamma} $ is a smooth function of energy $E_{\gamma}$, $f(E_{\gamma})$ ($V_{\gamma \gamma}$ almost do not fluctuate) \cite{S2008Natur.452..854R,S1995chao.dyn.11001S}. In this case, we end up with $F(t\rightarrow \infty) \sim \sum_{\alpha} c_{\alpha}^* b_{\alpha} |V_{\alpha \alpha}|^3$ up to some residue due to finite-size and conservation laws \cite{SPhysRevLett.123.010601}. (We assume $V=W$ for simplicity of the argument.) We can state,
\begin{eqnarray}
\text{Tr}(V \mathbb{I}) \sim \text{Tr}(V \Ket{\psi(0)}\Bra{\psi(0)}) = \Bra{\psi(0)}V \Ket{\psi(0)}
\end{eqnarray}
because $\Ket{\psi(0)}$ is drawn from a Haar-distributed ensemble (detailed in the next paragraph). Then, under the assumption of $\text{Tr}\left(V\right)=0$, $\Bra{\psi(0)}V \Ket{\psi(0)} = \sum_{\alpha} c_{\alpha}^* b_{\alpha} = 0$. Since $V_{\gamma \gamma}$ do not fluctuate significantly via ETH's second criteria \cite{S2008Natur.452..854R} and in fact the support of distribution of $V_{\gamma \gamma}$ shrinks around the microcanonical ensemble value in the thermodynamic limit if we assume the strong form of ETH \cite{SPhysRevLett.105.250401}, $F(t\rightarrow \infty) \rightarrow 0$ for chaotic spin systems. However, in order to extract the rate of decaying to zero  e.g. power-law in chaotic spin systems with conserved quantities, and/or the residue due to finite-size, one needs a more rigorous analysis where the fluctuations of the diagonal elements $V_{\gamma \gamma}$ around the smooth function $f(E)$ are included in the second assumption of ETH \cite{SPhysRevLett.123.010601}.

$\text{Tr}V \sim \Braket{\psi|V|\psi}$ holds for a pure state $\Ket{\psi}$ randomly drawn  from uniform distribution induced by Haar measure \cite{S2017AnP...52900350L}. These so-called Haar-distributed random states, unlike random product states, are close to maximally entangled states \cite{S2006CMaPh.265...95H}. By being close to maximally-entangled states, Haar-distributed random pure states behave as typical states on which \emph{canonical typicality} \cite{Spopescu2006entanglement,SPhysRevLett.96.050403} could emerge. Canonical typicality is defined as the following: Consider a system $U$ with Hilbert space dimension $d_u$, composed of a subsystem $S$ with dimension $d_s$ and its environment $E$ of dimension $d_e$ which is significantly larger than the system itself $d_e \gg d_s$. In a system with equiprobable states, the state of the system would be $\rho_u = \mathbb{I}/d_u$. Hence the subsystem would be in a canonical state, $\Omega_s = \text{Tr}_e \rho_u$. Now if we take a pure typical state for the system $\Ket{\psi}$ instead of $\rho_u$, the subsystem state can be written as $\rho_s = \text{Tr}_e\Ket{\psi}\Bra{\psi}$. Canonical typicality asserts that $\rho_s \sim \Omega_s$ for typical states $\Ket{\psi}$. This means that typical system state $\Ket{\psi}$ is locally indistinguishable from $\rho_u$. As stated in Ref. \cite{Spopescu2006entanglement}, canonical typicality emerges because of `massive entanglement between the subsystem and the environment' which is a feature of typical states. These ideas are established in Ref. \cite{SPhysRevLett.96.050403} and more generally in Ref. \cite{Spopescu2006entanglement} via invoking Levy's lemma \cite{S2017AnP...52900350L} for systems with constraint of energy conservation. Later studies showed that energy constraint on the system is not required for canonical typicality to emerge \cite{SPhysRevLett.99.160404} and it is possible to formulate the principle for mechanical observables \cite{SPhysRevLett.99.160404,SPhysRevLett.108.240401}. Canonical typicality for a mechanical observable reads \cite{SPhysRevLett.108.240401,S2017AnP...52900350L} 
\begin{eqnarray}
P \left[\bigg\vert \text{Tr} V - \Braket{\psi |V| \psi} \bigg\vert \geq \epsilon \right] \lesssim \text{exp} \left(- \mathcal{O}(d_u)\right),
\end{eqnarray}
where $\epsilon$ is a small parameter. Hence, the probability that $\Braket{\psi |V| \psi}$ deviates from $\text{Tr} V$ decreases exponentially in the system size. Therefore, for big enough many-body systems Haar-distributed states could very well imitate an equiprobable state, $\rho_u=\mathbb{I}$ in the calculation of observable expectation values. This theory has been used in numerical studies to compute OTOCs at infinite-temperature \cite{SPhysRevB.96.020406,S2018arXiv180711085D}.

\begin{figure}
\centering
\subfloat[]{\label{supfig:fig2a}\includegraphics[width=0.24\textwidth]{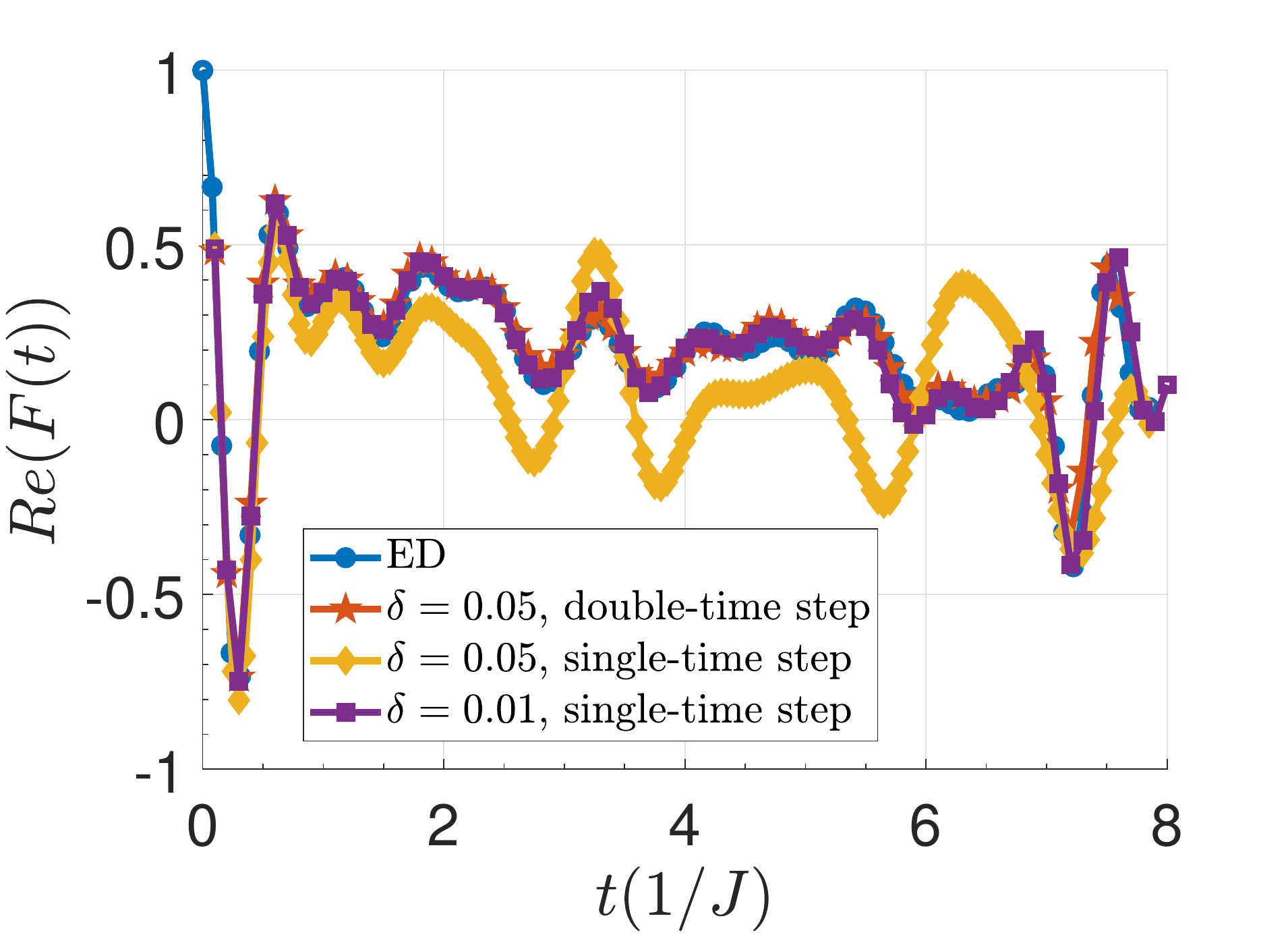}}\hfill \subfloat[]{\label{supfig:fig2b}%
\includegraphics[width=0.24\textwidth]{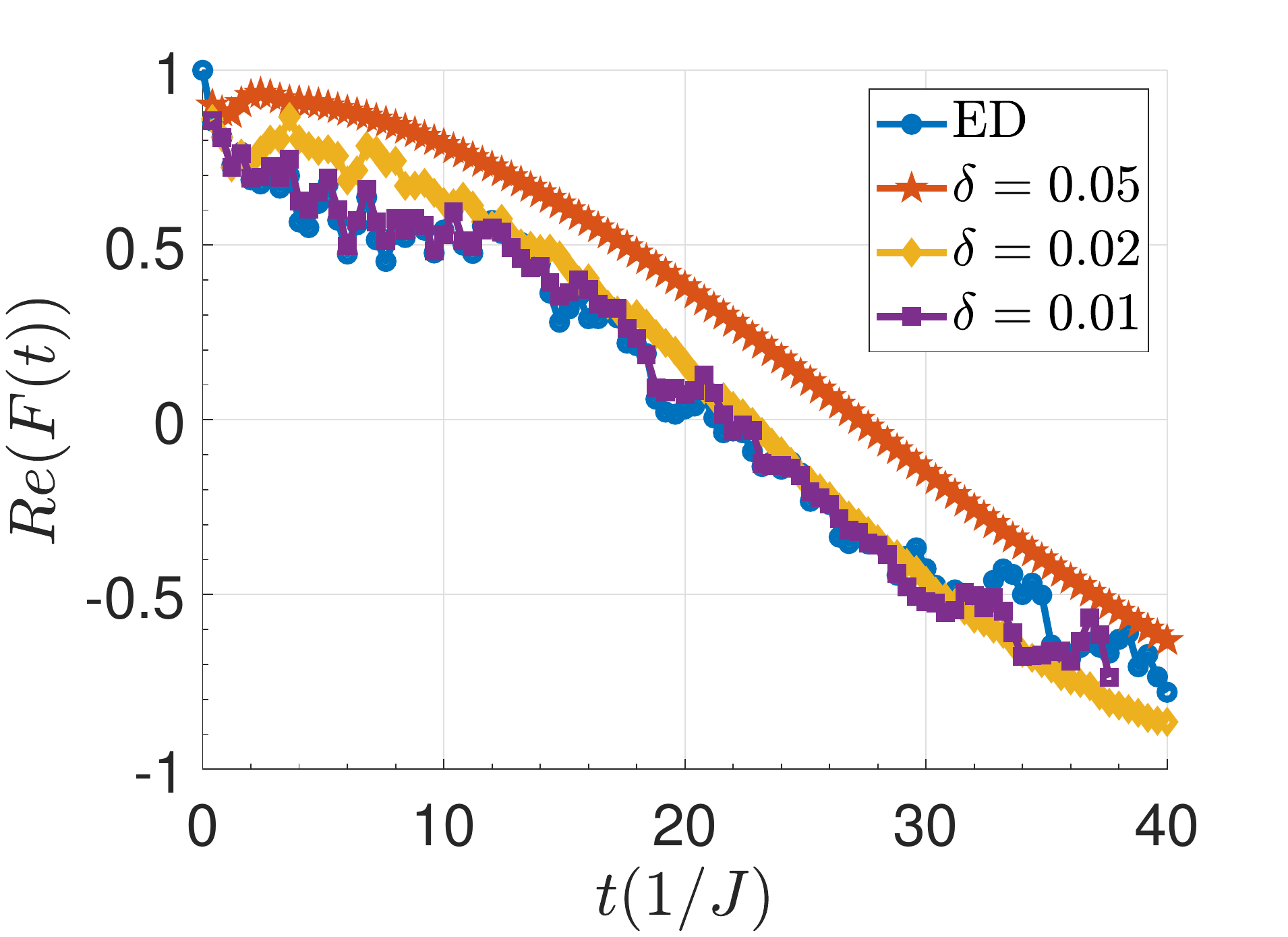}}
\hfill 
\caption{Real-time dynamics of OTOC at (a) $J_z/J=0.5$ and (b) $J_z/J=4$ at $N=14$. (a) Blue-circles: exact diagonalization, red-pentagons: double-time steps with $\delta=0.05$, yellow-diamonds: single-time steps with $\delta=0.05$ and purple-squares: single-time steps with $\delta=0.01$. (b) Blue-circles: exact diagonalization, red-pentagons: double-time steps with $\delta=0.05$, yellow-diamonds: double-time steps with $\delta=0.02$ and purple-squares: double-time steps with $\delta=0.01$. Both subfigures are at $h/J=0$.}
\label{SuppFig2}
\end{figure}

\section{Density-matrix renormalization group computations}

In order to compute the real-time dynamics of OTOC with MPS, we first find the ground state of the system with DMRG where we limit our computation to maximum 10 sweeps and set it to initial state. Then we time-evolve the initial state approximately \cite{SPhysRevB.91.165112}. Note that this approach is accurate as long as the time increment $\delta$ is small enough. Fig.~\ref{supfig:fig2a} compares three different methods in the calculation of the OTOC at $J_z/J=0.5$ (XY-phase). While ED stands for exact diagonalization, single-time step means using only real time step $\delta$ and double-time means utilizing two complex time steps at a real time step in order to decrease the scaling of the error \cite{SPhysRevB.91.165112}. Fig.~\ref{supfig:fig2b} shows different $\delta$ values with double-time steps in the anti-ferromagnetic phase of the XXZ model. Due to our benchmarking results, we set $\delta=0.01$ for our time evolution computations. For the calculation of the ground state contribution to OTOC as the leading-order term in the Ising-ordered phases, we compute the first three states with lowest energies via DMRG.

\bibliographystyle{apsrev4-1}
%\bibliography{SBibliography} % The references (bibliography) information are stored in the file named 


\end{document}